\documentclass[amsfonts,amsmath,amssymb,aps,pra,showpacs,showkeys,preprint]{revtex4-1}

\usepackage{graphicx}
\usepackage[english]{babel}
\usepackage{bm}
\begin{document}
\title{The PVLAS experiment: measuring vacuum magnetic birefringence and dichroism with a birefringent Fabry-Perot cavity}

\author{F. \surname{Della Valle},\email{federico.dellavalle@ts.infn.it} E. Milotti}
\affiliation{INFN, Sez. di Trieste and Dip. di Fisica, Universit\`a di Trieste, via A. Valerio 2, I-34127 Trieste, Italy}
\author{A. Ejlli, U. Gastaldi, G. Messineo, G. Zavattini}
\affiliation{INFN, Sez. di Ferrara and Dip. di Fisica e Scienze della Terra, Universit\`a di Ferrara, via Saragat 1, Edificio C, I-44122 Ferrara, Italy}
\author{R. Pengo, G. Ruoso}
\affiliation{INFN, Lab. Naz. di Legnaro, viale dell'Universit\`a 2, I-35020 Legnaro, Italy}


\pacs{07.60.Fs,14.80.Va,42.50Xa,78.20Ls}
\keywords{Quantum Electrodynamics, Magnetic birefringence, Polarimetry, Axions, Milli-charged particles}
\date{Received: date / Accepted: date}

\begin{abstract}
Vacuum magnetic birefringence was predicted long time ago and is still lacking a direct experimental confirmation. Several experimental efforts are striving to reach this goal, and the sequence of results promises a success in the next few years. This measurement generally is accompanied by the search for hypothetical light particles that couple to two photons. The PVLAS experiment employs a sensitive polarimeter based on a high finesse Fabry-Perot cavity. In this paper we report on the latest experimental results of this experiment. The data are analysed taking into account the intrinsic birefringence of the dielectric mirrors of the cavity. Besides the limit on the vacuum magnetic birefringence, the measurements also allow the model-independent exclusion of new regions in the parameter space of axion-like and milli-charged particles. In particular, these last limits hold also for all types of neutrinos, resulting in a laboratory limit on their charge.
\end{abstract}

\maketitle

\section{Introduction}	

Vacuum magnetic birefringence is a very small macroscopic quantum effect stemming from the 1936 Euler-Heisenberg-Weisskopf effective Lagrangian density for slowly varying electromagnetic fields \cite{EHW} that, to lowest order, reads:
\begin{equation}
{\cal L}_{\rm EHW} = \frac{1}{2\mu_{\rm 0}}\left(\frac{E^{2}}{c^{2}}-B^{2}\right)+\frac{A_{e}}{\mu_{\rm 0}}\left[\left(\frac{E^{2}}{c^{2}}-B^{2}\right)^{2}+7\left(\frac{\vec{E}}{c}\cdot\vec{B}\right)^{2}\right].
\label{LEH}
\end{equation}
Here 
\begin{equation}
A_{e} = \frac{2}{45\mu_{0}}\frac{\alpha^{2}\mathchar'26\mkern-10mu\lambda_e^{3}}{m_{e}c^{2}} = 1.32\times10^{-24}~{\rm T}^{-2},
\label{Ae}
\end{equation}
$\mathchar'26\mkern-10mu\lambda_e=\hbar/m_ec$ being the Compton wavelength of the electron, $\alpha={e^2}/{(4\pi\varepsilon_0\hbar c)}$ the fine structure constant, and $m_e$ the electron mass.
The first term in Equation (\ref{LEH}), quadratic in the fields, is the classical Lagrangian corresponding to Max\-well's equations in vacuum, for which the superposition principle holds and no light-by-light interaction is expected. The other terms, instead, imply that Electrodynamics is nonlinear even in vacuum, giving rise to a new class of observable effects.

\begin{figure}[htb]
\begin{center}
\includegraphics[width=10cm]{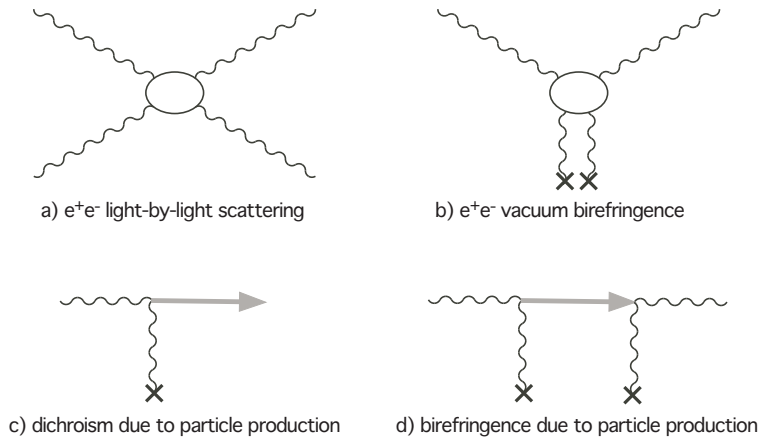}
\end{center}
\caption{Lowest order elementary processes leading to magnetic birefringence and dichroism.}
\label{4photon}
\end{figure}

The Quantum Electrodynamics (QED) representation of the simplest phenomena we are interested in is given by the Feynman diagrams shown in Figures \ref{4photon}a) and \ref{4photon}b), in which four photons interact through a virtual $e^+e^-$ pair. In the \ref{4photon}b) diagram, two photons interact with an external field; this is the process that leads, in vacuum, to magnetic birefringence, namely to different indices of refraction for light polarised parallel and perpendicular to an external magnetic field $B_{\rm ext}$. Considering the complex index of refraction $n+i\kappa$, it can be shown \cite{Delta_n_QED,Adler1971} that the magnetic birefringence derived from Equation (\ref{LEH}) is
\begin{equation}
\Delta n^{\rm (EHW)}=n_{\parallel}^{\rm (EHW)}-n_{\perp}^{\rm (EHW)} = 3 A_{e} B_{\rm ext}^2.
\label{birifQED}
\end{equation}
This corresponds to
\begin{equation}
\Delta n^{\rm (EHW)}=2.5\times10^{-23}\qquad @~B_{\rm ext}=2.5{\rm~T}.
\label{DeltanQED}
\end{equation}
The calculations also show that the magnetic dichroism is instead negligible \cite{Adler1971}: no appreciable imaginary part $\kappa$ of the index of refraction is predicted.

Magnetic birefringence accompanied by magnetic dichroism could, though, be generated in vacuum through the creation of so far hypothetical light bosonic spin-zero axion-like particles (ALPs) \cite{Maiani1986}, in an analog of the Primakoff effect \cite{Primakov1951}. The two processes generating dichroism and birefringence are shown, respectively, in Figures \ref{4photon}c) and \ref{4photon}d). Two different Lagrangians describe the pseudoscalar and the scalar cases:
\[
{\cal L}_{a} = g_a\phi_{a} \vec{E}\cdot \vec{B} \quad{\rm and}\quad
{\cal L}_{s} = g_s\phi_{s} \left(E^{2}-B^{2} \right),
\]
where $g_a$ and $g_s$ are the coupling constants of a pseudoscalar field $\phi_a$ and of a scalar field $\phi_s$, respectively, and the natural Heaviside-Lorentz units are used, so that 1~T $=\sqrt{\frac{\hbar^{3}c^{3}}{e^{4}\mu_{0}}}= 195$~eV$^2$ and 1~m $=\frac{e}{\hbar c}=5.06\times10^{6}$~eV$^{-1}$. One finds \cite{Raffelt1988}
\begin{eqnarray}
\nonumber
&|\Delta n^{\rm (ALP)}|=\displaystyle n_{\parallel}^{a}-1 = n_{\perp}^{s}-1 = \frac{g_{a,s}^2B_{\rm ext}^{2}}{2m_{a,s}^{2}}\left(1-\frac{\sin2x}{2x}\right),\\
&|\Delta\kappa^{\rm (ALP)}|=\displaystyle\kappa_{\parallel}^{a} = \kappa_{\perp}^{s} = \frac{2}{\omega L}\left(\frac{g_{a,s}B_{\rm ext}L}{4}\right)^{2}\left(\frac{\sin x}{x}\right)^{2},
\label{ALPs}
\end{eqnarray}
where $m_{a,s}$ are the masses of the particles, $x=\frac{Lm_{a,s}^{2}}{4\omega}$ in vacuum, $\omega$ is the photon energy, and $L$ is the magnetic field length.

Consider now the vacuum fluctuations of particles with charge $\pm\epsilon e$ and mass $m_{\epsilon}$ as discussed in Reference \cite{Ahlers2007}. The photons traversing a uniform magnetic field may interact with such fluctuations, resulting in a phase delay and, if the photon energy $\hbar\omega>2m_\epsilon c^2$, in a pair production. We consider separately the case of Dirac fermions (Df) and of scalar (sc) bosons. The indices of refraction of photons with polarisation respectively parallel and perpendicular to the external magnetic field have two different mass regimes defined by a dimensionless parameter $\chi$:
\begin{equation}
\chi\equiv\frac{3}{2}\frac{\hbar\omega}{m_{\epsilon}c^{2}}\frac{\epsilon e B_{\rm ext}\hbar}{m_{\epsilon}^{2}c^{2}}.
\label{chi}
\end{equation}

In the case of fermions, it can be shown that \cite{Tsai1975,Ahlers2007}
\[
\Delta n^{\rm (Df)}=A_{\epsilon} B_{\rm ext}^{2}
\left\{\begin{array}{ll}
3 & \textrm { for  } \chi \ll 1 \\
\displaystyle-\frac{9}{7}\frac{45}{2}\frac{\pi^{1/2}2^{1/3}\left[\Gamma\left(\frac{2}{3}\right)\right]^{2}}{\Gamma\left(\frac{1}{6}\right)}\chi^{-4/3} & \textrm{ for   }\chi\gg 1
\end{array}\right.
\]
where \[A_{\epsilon}=\frac{2}{45\mu_{0}}\frac{\epsilon^{4}\alpha^{2} \mathchar'26\mkern-10mu\lambda_\epsilon^{3}}{m_{\epsilon}c^{2}}\] in analogy to Equation (\ref{Ae}). In the limit of large masses ($\chi\ll1$) the expression reduces to Equation (\ref{birifQED}) with the substitution of $\epsilon e$ with $e$ and $m_{\epsilon}$ with $m_{e}$. Note that for small masses ($\chi\gg1$) the birefringence depends on the parameter $\chi^{-4/3}$ resulting in a net dependence of $\Delta n^{\rm (Df)}$ with $B_{\rm ext}^{2/3}$ rather than $B_{\rm ext}^{2}$ as in Equation (\ref{birifQED}). For dichroism one finds \cite{Tsai1974,Ahlers2007}
\[
\Delta \kappa^{\rm (Df)}=\frac{1}{8\pi}\frac{\epsilon^3e\alpha\lambda B_{\rm ext}}{m_\epsilon c}
\left\{\begin{array}{ll}
\sqrt{\frac{3}{32}}\,e^{-4/\chi} & \textrm { for  } \chi \ll 1 \\
\displaystyle\frac{2\pi}{3\,\Gamma(\frac{1}{6})\Gamma(\frac{13}{6})}\,\chi^{-1/3} & \textrm{ for   }\chi\gg 1.
\end{array}\right.
\]

The results for the case of milli-charged scalar particles are very similar to the case of Dirac fermion case \cite{Ahlers2007}. Again there are two mass regimes defined by the same parameter $\chi$ of expression (\ref{chi}). In this case the magnetic birefringence is
\[
\Delta n^{\rm (sc)}=A_{\epsilon} B_{\rm ext}^{2}\left\{\begin{array}{ll}
\displaystyle-\frac{6}{4} & \textrm{ for  } \chi \ll 1 \\
\displaystyle\frac{9}{14}\frac{45}{2}\frac{\pi^{1/2}2^{1/3}\left[\Gamma\left(\frac{2}{3}\right)\right]^{2}}{\Gamma\left(\frac{1}{6}\right)}\chi^{-4/3} & \textrm{ for   }\chi\gg 1.
\end{array}\right.\nonumber
\]
The dichroism is given by
\[
\Delta \kappa^{\rm (sc)}=\frac{1}{8\pi}\frac{\epsilon^3e\alpha\lambda B_{\rm ext}}{m_\epsilon c}
\left\{\begin{array}{ll}
-\sqrt{\frac{3}{8}}\,e^{-4/\chi} & \textrm { for  } \chi \ll 1 \\
\displaystyle-\frac{\pi}{3\,\Gamma(\frac{1}{6})\Gamma(\frac{13}{6})}\,\chi^{-1/3} & \textrm{ for   }\chi\gg 1.
\end{array}\right.
\]
As can be seen, there is a sign difference with respect to the case of Dirac fermions, both for birefringence and for dichroism.


The PVLAS (Polarisation of Vacuum with LASer) experiment in Ferrara is the fourth generation of a measurement scheme that dates back to the end of the '70s \cite{Iacopini1979}. Previous experimental efforts were based at CERN \cite{Iacopini1981}, at BNL \cite {Cameron1993}, and at Legnaro (Italy) \cite{Bregant2008}. The experiment aims at the direct measurement of the small polarisation changes undergone by a linearly polarised laser beam traversing a dipole magnetic field in vacuum. To this end, a pair of polarising prisms, two permanent magnets, an optical high-finesse Fabry-Perot cavity, and heterodyne detection are employed. A quarter-wave-plate placed after the Fabry-Perot switches the measurement from ellipticity to rotation (dichroism). The signal is detected in the extinguished beam with polarisation orthogonal to the input polarisation.

The Fabry-Perot cavity has the role of lengthening the optical path inside the magnetic field. It is realised with two dielectric mirrors with extremely high reflectivity. Unfortunately, the mirrors have a small intrinsic linear birefringence in reflection. A first consequence of this fact is that, if linearly polarised laser light is at maximum resonance inside the cavity, the orthogonal polarisation component is not. This means that the amplitude of the observed signal is reduced; this fact comes out evident by calibrating the ellipsometer with magnetic birefringence in gas (Cotton-Mouton -- or Voigt -- effect) \cite{Rizzo1997}. Recent anomalously low Cotton-Mouton results could perhaps be explained in this way \cite{Mei2010}. As a second consequence, ellipticities and rotations are mixed, due to the mirrors' birefringence. As we will see, both phenomena can be managed, in some cases even with profit. Moreover, the intrinsic birefringence of the mirrors may play a role in the excess noise currently observed in the PVLAS experiment.

In this article we present a detailed account of the polarimetric method employed by the PVLAS experiment, with a novel interpretation of the experimental data. What we describe here has consequences for all the experiments that use Fabry-Perot cavities for polarimetry, and in particular for those trying to measure vacuum magnetic birefringence. Section \ref{conti} analyses the experimental scheme, taking into account the intrinsic birefringence of the mirrors. Section \ref{ExpSect} describes the experimental set-up with the calibration measurements. Then the measurement of the mirrors' equivalent wave-plates and of the two resonance curves are presented. In Section \ref{Misure} the ellipticity and rotation measurements in vacuum are discussed, together with the new limits on the existence of axion-like and milli-charged particles.

\section{The PVLAS experimental method}
\label{conti}

\begin{figure}[htb]
\begin{center}
\includegraphics[width=10cm]{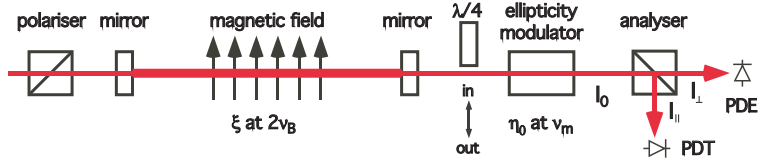}
\end{center}
\caption{Scheme of the PVLAS polarimeter. PDE: Extinction Photodiode; PDT: Transmission Photodiode.}
\label{Scheme}
\end{figure}

In Figure \ref{Scheme}, a scheme of the PVLAS polarimeter is shown. Linearly polarised light (wavelength $\lambda$) is fed to a Fabry-Perot optical cavity. The cavity beam traverses the bore of a dipole magnet, with the magnetic field making an angle $\phi(t)$, variable in time, with respect to the polarisation direction. A variable ellipticity $\eta(t)$ is added to the polarisation of the beam transmitted by the cavity. For rotation measurement, a quarter-wave-plate ($\lambda/4$) is inserted at the exit of the cavity with one of its axes aligned to the input polarisation, transforming the rotation eventually acquired by the beam inside the magnetic field region into an ellipticity (and, at the same time, the ellipticity into a rotation). Finally a polariser, crossed with respect to the input prism, extinguishes the polarisation component of the beam parallel to the input polarisation. The residual intensity is then collected with a light detector and Fourier analysed.

\begin{figure}[htb]
\begin{center}
\includegraphics[width=4cm]{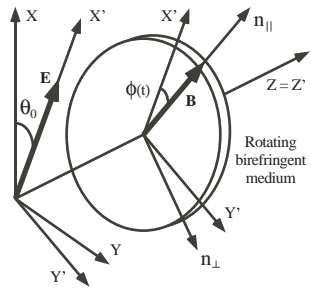}
\end{center}
\caption{Reference frame for the calculations. $XY$: laboratory coordinates; $X'$: direction of the electric field as defined by the polariser; $n_\parallel$: direction of the magnetic field, rotating around the beam path $Z$ at a frequency $\nu_B$.}
\label{frame}
\end{figure}

In order to calculate the effect, we use Jones' matrices \cite {Jones1948} to describe the beam and the optical elements. 
The most general optical element describing linear magnetic birefringence and dichroism can be written, in its own axes and neglecting an overall attenuation factor, as
\[
\mathbf{X}_0=\left(\begin{array}{cc}e^{\xi} & 0 \\0 & 1\end{array}\right),
\]
where $\xi$ is a small complex number that we write as $\xi=i\,2\psi-2\theta$. Here $2\psi$ is the phase difference between the two polarisation directions added by the optical element and $1-e^{-2\theta}$ is the fraction of the absorbed electric field. Without loss of generality, the $x$ direction ($X'$ direction of Figure \ref{frame}) is considered as the absorbing as well as the slow axis. The value $\psi$ is the maximum ellipticity\footnote{Ellipticity is the ratio of the minor to the major axis of the ellipse described by the electric field vector of light.} that light can acquire due to $\mathbf{X}_0$, while $\theta$ is the maximum rotation. In the case of the vacuum birefringence of Equation (\ref{birifQED}), the ellipticity $\psi$ for a length $L=1.6$~m of a magnetic field $B_{\rm ext}=2.5$~T and light wavelength $\lambda=1~\mu$m is
\begin{equation}
\psi_{\rm QED}=\pi\,\frac{\Delta n^{\rm(EHW)}L}{\lambda}=1.2\;10^{-16}.
\label{ell_QED}
\end{equation}
Placing $\mathbf{X}_0$ at an angle $\phi$ with respect to the polarisation direction, one finds
\[
\mathbf{X}(\phi)=\frac{1}{2}\left(\begin{array}{cc}1-\cos2\phi+e^\xi (1+\cos2\phi)&-\left(1-e^\xi\right)\sin2\phi\\
-\left(1-e^\xi\right)\sin2\phi&1+\cos2\phi+e^\xi(1-\cos2\phi)\end{array}\right).
\]

To show the salient features of our polarimetric method, we begin with neglecting the effect of the Fabry-Perot cavity. The electric field after the analyser is then represented by
\[
\mathbf{E}(\phi)=E_0
\left(\begin{array}{cc}0&0\\0&1\end{array}\right)
\cdot
\left(\begin{array}{cc}1&i\,\eta\\i\,\eta&1\end{array}\right)
\cdot
\left(\begin{array}{cc}q&0\\0 &q^*\end{array}\right)
\cdot
\mathbf{X}(\phi)
\cdot
\left(\begin{array}{c}1\\0\end{array}\right).
\]
In this formula, from left to right, one finds the Jones matrices of the analyser $\mathbf{A}$, of the ellipticity modulator $\mathbf{H}$ ($\theta,\psi\ll\eta\ll1$), and of the quarter-wave-plate $\mathbf{Q}$. In this last matrix, $q=1$ for ellipticity measurements, when the wave-plate is out of the optical path and $\mathbf{Q}$ therefore coincides with the identity matrix $\mathbf{I}$, whereas $q=(1+i)/\sqrt{2}$ for rotation measurements. For ellipticity measurements (quarter-wave-plate not inserted), the intensity collected at the photodiode PDE is
\begin{equation}
I_\perp^{\rm ell}(\phi)=I_0\,\left(\eta^2+2\eta\psi\sin2\phi\right)+\rm{higher~order~terms.}
\label{ell0}
\end{equation}
For rotation measurements, with the quarter-wave-plate inserted,
\begin{equation}
I_\perp^{\rm rot}(\phi)=I_0\left(\eta^2+2\eta\theta\sin2\phi\right)+\rm{higher~order~terms.}
\label{rot0}
\end{equation}
The light having the same polarisation as the input is collected at the photodiode PDT and has intensity \[I_\parallel\approx I_0=\varepsilon_0c\frac{E_0^2}{2}.\]

The heterodyne method is employed to measure $\psi$ and $\theta$: the angle $\phi$ is varied linearly in time as $\phi(t)=2\pi\nu_Bt+\phi_B$, and $\eta$ as $\eta(t)=\eta_0\cos(2\pi\nu_m t+\phi_m)$, with $\nu_B\ll\nu_m$. The sought for value of each of the quantities $\psi$ and $\theta$ can be extracted from the measurement of $I_\parallel$ and from the amplitude and phase of three components in a Fourier transform of the extinguished intensity: the component $I_{2\nu_m}$ at $2\nu_m$ and the components $I_\pm$ at $\nu_m\pm2\nu_B$.
By using a lock-in amplifier to demodulate the residual intensity at the frequency $\nu_m$, instead of $I_+$ and $I_-$ there is a single component at $2\nu_B$, and the resulting ellipticity and rotation signals are
\begin{equation}
\psi,\theta=\frac{I_{2\nu_B}}{2\sqrt{2\,I_0\,I_{2\nu_m}}}=\frac{I_{2\nu_B}}{I_{2\nu_m}}\frac{\eta_0}4.
\label{solvingLockIn}
\end{equation}
The ellipticity and rotation signals come with a well defined phase $2\phi_B$. With reference to Figure \ref{frame}, one can see that the value of $\phi_B$ is $-\theta_0$, with $\theta_0$ the angle between a reference direction $X$ and the polarisation direction. With this position, the axes of $\mathbf{X}_0$ coincide with the laboratory axes $(XY)$ and the ellipticity is a maximum at the time $t_0=(\theta_0+\pi/4)/(2\pi\nu_B)$. We will return to this topic in the calibration section.

In the absence of signals due to magnetic birefringence or dichroism, the noise level at the signal frequency translates into an upper limit for the measured quantity.

\subsection{The Fabry-Perot cavity as an optical path multiplier}
To take into account the multiple reflections of the Fabry-Perot cavity, we consider the physical parameters of the mirrors, namely the reflectivity, transmissivity, and losses, $R$, $T$, and $P$ (assumed equal for both mirrors), such that $R+T+P=1$. If $d$ is the distance between the two mirrors, let $\delta=4\pi d/\lambda$ be the phase acquired by the light in a round trip. Then one can write, for the electric field after the cavity,
\begin{eqnarray}
\nonumber
\mathbf{E}_{\rm out}(\delta,\phi)&=&\left(\begin{array}{c}E_{{\rm out},\parallel}\\E_{{\rm out},\perp}\end{array}\right)=E_0\,\sum_{n=0}^\infty\left[Re^{i\delta}\,\mathbf{X}^2(\phi)\right]^n\cdot Te^{i\delta/2}\,\mathbf{X}(\phi)\cdot\left(\begin{array}{c}1\\0\end{array}\right)\\
&=&E_0\,\left[\mathbf{I}-Re^{i\delta}\,\mathbf{X}^2(\phi)\right]^{-1}\cdot Te^{i\delta/2}\,\mathbf{X}(\phi)\cdot\left(\begin{array}{c}1\\0\end{array}\right),
\label{FP}
\end{eqnarray}
and for the electric field after the analyser
\begin{equation}
\mathbf{E}(\delta,\phi)=\mathbf{A}\cdot\mathbf{H}\cdot\mathbf{Q}\cdot\mathbf{E}_{\rm out}(\delta,\phi).
\label{DetectedField}
\end{equation}

In the case of ellipticity measurements, since at resonance $\delta=0$ (mod $2\pi$), and given that $R\approx1$, the intensity collected by photodiode PDE, at the lowest order, is
\begin{equation}
I_\perp^{\rm ell}(\phi)\simeq I_0\,\left[\eta^2+\frac{4\eta\psi}{1-R}\sin2\phi\right].
\label{ellFP}
\end{equation}
Analogously, in the case of rotation measurements, one has
\begin{equation}
I_\perp^{\rm rot}(\phi)\simeq I_0\,\left[\eta^2+\frac{4\eta\theta}{1-R}\sin2\phi\right],
\label{rotFP}
\end{equation}
while
\begin{equation}
I_\parallel\approx I_0=\varepsilon_0c\frac{E_0^2}{2}\,\frac{T^2}{(T+P)^2}.
\label{transmissionFP}
\end{equation}
By comparing these formulas with the corresponding ones calculated above without the Fabry-Perot cavity [Equations (\ref{ell0}) and (\ref{rot0})], one sees that the expressions are very similar, with the latter ones having the signals $\psi$ and $\theta$ of Equation (\ref{solvingLockIn}) amplified by a factor
\[
N=\frac{2}{1-R}\approx\frac{2{\cal F}}{\pi},
\]
where ${\cal F}$ is the {\em finesse} of the cavity, that can be up to $\sim10^6$ \cite{DellaValle2014OE}. This can be interpreted as a {\em lengthening} of the optical path by a factor $N$, as the very form of Equation (\ref{FP}) suggests. Besides heterodyne detection, high amplification is another key feature of the polarimetric technique adopted by the PVLAS experiment. In this way, the ellipticity of Equation (\ref{ell_QED}) becomes of order $10^{-10}$.

We now introduce another issue of the Fabry-Perot cavity that will be fully discussed in the next paragraph. Let us suppose that the condition $\delta=0$ (mod $2\pi$) is not fully matched, namely that the Fabry-Perot cavity is not exactly locked to the top of the resonance curve. The two Equations (\ref{ellFP}) and (\ref{rotFP}) become, respectively,
\begin{equation}
I_\perp^{\rm ell}(\phi)\simeq I_0\left[\eta^2+\eta\,\frac{2N\psi-N^2\theta\sin\delta}{1+N^2\sin^2(\delta/2)}\,\sin2\phi\right]
\label{CrossTalk1}
\end{equation}
for the case of ellipticity measurements, and
\begin{equation}
I_\perp^{\rm rot}(\phi)\simeq I_0\left[\eta^2+\eta\,\frac{2N\theta+N^2\psi\sin\delta}{1+N^2\sin^2(\delta/2)}\,\sin2\phi\right]
\label{CrossTalk2}
\end{equation}
for rotation measurements. Equation (\ref{transmissionFP}) becomes instead \[I_\parallel\approx I_0=\varepsilon_0c\frac{E_0^2}{2}\,\frac{T^2N^2/4}{1+N^2\sin^2(\delta/2)}.\] One can see that, in a cavity locked at $\delta\neq0$, there is a cross talk between the birefringence and dichroism signals as defined by Equation (\ref{solvingLockIn}): a rotation is measured even in the case $\psi\neq0$ and $\theta=0$. Conversely, in the case $\psi=0$ and $\theta\neq0$, a signal mimicking a birefringence is observed. 

\subsection{Mirror birefringence}
\label{MirrorBirefringence}
Let us now tackle the problem of dealing with birefringent mirrors \cite{Zavattini2006}. If $\alpha_{1,2}$ are the small phase differences acquired by light in just one reflection by the mirrors, one must introduce in the above calculations the wave-plates
\[
\mathbf{M}_{1,2}=\left(\begin{array}{cc}e^{i\,\alpha_{1,2}/2}&0\\0&e^{-i\,\alpha_{1,2}/2}\end{array}\right),
\]
where both $\alpha$'s can be thought of as positive quantities, without loss of generality. Assuming, for simplicity (see Section \ref{MirrorStudies} for the more general case) that the slow axes of the mirror wave-plates are both aligned to the input polarisation, the polarisation auto-states of the Faby-Perot cavity are given by
\[
\left(\begin{array}{c}\left[1-R\,e^{i[\delta+(\alpha_1+\alpha_2)/2]}\right]^
{-1}\\
0\end{array}\right){\rm~and~}
\left(\begin{array}{c}0\\
\left[1-R\,e^{i[\delta-(\alpha_1+\alpha_2)/2]}\right]^{-1}\end{array}\right).
\]
The above equations show that the resonance curves of the two polarisation modes are no longer centred at $\delta=0$, and are separated by the quantity
\[
\alpha=\alpha_1+\alpha_2.
\]
In other words, the two polarisations cannot resonate at the same time inside the cavity.

In the PVLAS experiment, the emission frequency of the laser is locked to the resonance frequency of the cavity by means of a feedback electronic circuit based on the Pound and Drever locking scheme, in which the error signal is carried by the light reflected from the cavity through the input polariser. As a consequence, while the light having the input polarisation is at the top of the resonance curve ($\delta=-\alpha/2$), the orthogonal component is not. As the frequency width of the cavity is a few tens of hertz, for frequency differences of this order of magnitude the orthogonal component may be filtered significantly. Hence, as a first issue, when analysing the extinguished beam one has to necessarily take into account the fact that its intensity is reduced by the factor
\begin{equation}
k(\alpha)=\frac{1}{1+N^2\sin^2(\alpha/2)}\leq1.
\label{kappa}
\end{equation}
with respect to the other polarisation. By varying the input polarisation direction and the relative angular position of the two mirrors, it is possible to minimise the effect of the mirrors' wave-plates by aligning the slow axis of one mirror against the fast axis of the other. This ensures that the two curves are as near as possible, in which case $\alpha$ is equal to the difference $\Delta\alpha=\alpha_2-\alpha_1$.

\begin{figure}[htb]
\begin{center}
\includegraphics[width=10cm]{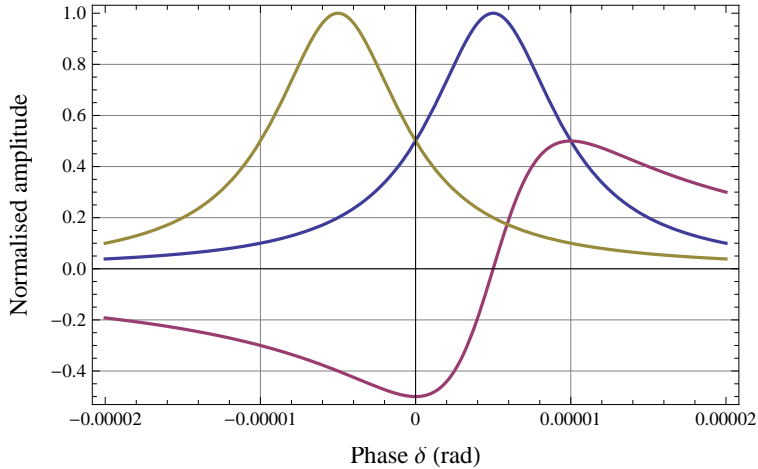}
\end{center}
\caption{From left to right: Transmitted intensity $I_\parallel$, amplitudes of the ellipticity and of the rotation signals of Equation (\ref{solvingLockIn}) in the case of a pure birefringence, as functions of the Fabry-Perot cavity phase $\delta$, for $\alpha=10^{-5}$ and $N=4\times10^5$. The Airy curves are normalised to unity; the rotation signal bears the same normalisation coefficient as the ellipticity. Transmitted intensity is centred at $\delta=-\alpha/2$, the other two curves at $\delta=\alpha/2$. The amplitude of the ellipticity signal at $\delta=-\alpha/2$ is a factor $k(\alpha)=0.2$ smaller than the maximum [see Equation (\ref{kappa})].}
\label{BirefringentAiry}
\end{figure}

As a second issue, analogously to Equations (\ref{CrossTalk1}) and (\ref{CrossTalk2}), a symmetrical mixing appears between rotations and ellipticities. In fact, the electric field at the exit of the cavity is
\[
\mathbf{E}_{\rm out}(\phi,\delta)=E_0\,\left[\mathbf{I}-Re^{i\delta}\,\mathbf{X}\cdot\mathbf{M}_1\cdot\mathbf{X}\cdot\mathbf{M}_2\right]^{-1}\cdot Te^{i\delta/2}\mathbf{X}\cdot\left(\begin{array}{c}1\\0\end{array}\right).
\]
From Equation (\ref{DetectedField}), the intensity at the detector for small $\alpha$'s, and $R\approx1$, is
\begin{equation}
I_\perp^{\rm ell}(\phi)=I_\parallel\left[\eta^2+\eta\,\frac{2N\psi-N^2\theta\left(\delta-\frac{\alpha}{2}\right)}{1+N^2\sin^2\left(\frac{\delta}{2}-\frac{\alpha}{4}\right)}\,\sin2\phi\right],
\label{EllipticityAiry}
\end{equation}
for the measurements of ellipticity, and
\begin{equation}
I_\perp^{\rm rot}(\phi)=I_\parallel\left[\eta^2+\eta\,\frac{2N\theta+N^2\psi\left(\delta-\frac{\alpha}{2}\right)}{1+N^2\sin^2\left(\frac{\delta}{2}-\frac{\alpha}{4}\right)}\,\sin2\phi\right]
\label{RotationAiry}
\end{equation}
for rotation measurements. Here
\[
I_\parallel=\varepsilon_0c\frac{E_0^2}{2}\,\frac{T^2N^2/4}{1+N^2\sin^2\left(\frac{\delta}{2}+\frac{\alpha}{2}\right)}.
\]
Note the similarity of the above equations with Equations (\ref{CrossTalk1}) and (\ref{CrossTalk2}). It can be shown that any small static ellipticity or rotation acquired before or after the cavity does not interfere with the signal at $2\nu_B$ and can thus be neglected. In Figure \ref{BirefringentAiry}, we plot the last three equations as functions of $\delta$ for the case $\theta=0$ (pure birefringence), for $N=4\times10^5$ and $\alpha=10^{-5}$.

If the laser is locked to the maximum value of $I_\parallel$ at $\delta=-\alpha/2$, one has, for an ellipticity measurement,
\begin{equation}
I_\perp^{\rm ell}(\phi)=I_\parallel\left[\eta^2+\eta k(\alpha)\,(2N\psi+N^2\theta\alpha)\,\sin2\phi\right],
\label{EllipticityMeasurement}
\end{equation}
while for a rotation measurement
\begin{equation}
I_\perp^{\rm rot}(\phi)=I_\parallel\left[\eta^2+\eta k(\alpha)\,(2N\theta-N^2\psi\alpha)\,\sin2\phi\right],
\label{RotationMeasurement}
\end{equation}
where $I_\parallel$ is given by Equation (\ref{transmissionFP}). With respect to Equations (\ref{ellFP}) and (\ref{rotFP}), the expected signals of ellipticity and rotation are attenuated by a factor $k(\alpha)$ [Equation (\ref{kappa})]. Moreover, a cross talk between the two measurement channels appears: even with $\theta=0$, a rotation $-kN^2\alpha\psi$ is observed. The ratio of the ``spurious" rotation and of the ``true" ellipticity is
\begin{equation}
R_{\theta,\psi}=-\frac{N}{2}\alpha,
\label{R}
\end{equation}
hence allowing a direct determination of the sum of the birefringences of the two mirrors. Analogously, even with $\psi=0$, an ellipticity $kN^2\alpha\theta$ appears.

In the absence of both signals, an upper limit coming from the measurement of one of the two quantities, ellipticity or rotation, translates in an upper limit also on the other one.

\subsection{Intrinsic noise of the polarimeter}

We now calculate the limit sensitivity of the apparatus. Starting from Equation (\ref{solvingLockIn}), if the noise at $\nu_m-2\nu_B$ is uncorrelated to the noise at $\nu_m+2\nu_B$, one must take into account a factor $\sqrt{2}$ due to the folding of the spectrum around $\nu_m$. If $I_{\rm noise}(2\nu_B)$ is the rms noise spectral density of the light intensity at the frequency of the signal, the expected peak sensitivity of the polarimeter is \[S_{2\nu_B}=\frac{I_{\rm noise}(2\nu_B)}{I_\parallel\eta_0}.\] Several intrinsic effects contribute to $S_{2\nu_B}$, all of which can be expressed as a noise in the light intensity impinging on the detector. We consider first the intrinsic rms shot noise due to the direct current $i_{\rm dc}$ in the detector \[i_{\rm shot}=\sqrt{2e\,i_{\rm dc}\,\Delta\nu}.\] According to Equations (\ref{ell0}) or (\ref{rot0}), the direct current inside the photodiode is given by $qI_\parallel\eta_0^2/2$, where $q$ is the efficiency of the detection process. However, any pair of crossed polarising prisms has a nonzero minimum extinction coefficient for intensity. For the best polarisers, the extinction coefficient can be as low as $\sigma^2\approx10^{-8}$. This effect introduces an additional term in the detected intensity which is written as $I_\parallel\sigma^2$. This leads to \[I_{\rm shot}=\sqrt{\frac{2e\,I_\parallel}{q}\left(\sigma^2+\frac{\eta_0^2}{2}\right)}\qquad{\rm and}\qquad S_{\rm shot}=\sqrt{\frac{2e}{qI_\parallel}\left(\frac{\sigma^2+\eta_0^2/2}{\eta_0^2}\right)}.\] Other effects contributing to the noise are the Johnson noise of the transimpedance $G$ of the photodiode \[I_J=\sqrt{\frac{4k_BT}{q^2G}},\qquad{\rm giving}\qquad S_J=\sqrt{\frac{4k_BT}{G}}\frac{1}{qI_\parallel\eta_0},\] the photodiode dark noise \[I_{\rm dark}=\frac{V_{\rm dark}}{qG},\qquad{\rm with}\qquad S_{\rm dark}=\frac{V_{\rm dark}}{G}\frac{1}{qI_\parallel\eta_0},\] and the relative intensity noise (RIN) of the light emerging from the cavity \[I_{\rm RIN}(\nu)=I_\parallel\,N_{\rm RIN}(\nu),\] giving \[S_{\rm RIN}(2\nu_B)=N_{\rm RIN}(\nu_m)\,\frac{\sqrt{(\sigma^2+\eta_0^2/2)^2+(\eta_0^2/2)^2}}{\eta_0,},\] where in the last equation we consider that the contributions of all the peaks in the Fourier spectrum add incoherently to the intensity noise at $\nu_m$, and that $\nu_B\ll\nu_m$.

\begin{figure}[htb]
\begin{center}
\includegraphics[width=10cm]{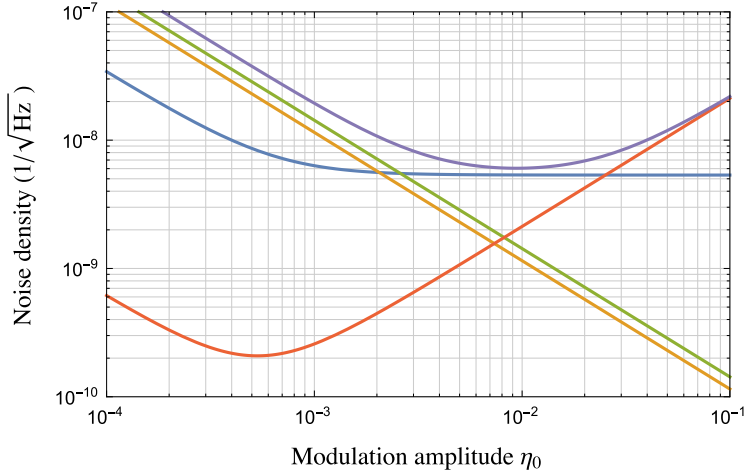}
\end{center}
\caption{Intrinsic peak noise of the polarimeter as a function of the ellipticity modulation amplitude $\eta_0$. On the left of the graph, from bottom upwards, one finds, in this order, RIN, shot, Johnson, and dark contributions, and the total intrinsic noise of the polarimeter.}
\label{IntrinsicNoise}
\end{figure}

Figure \ref{IntrinsicNoise} shows all the intrinsic contributions as functions of $\eta_0$ in typical operating conditions, with $q\approx0.7$~A/W, $I_\parallel=8$~mW, $\sigma^2=2\times10^{-7}$, $G=10^6~{\rm\Omega}$, $V_{\rm dark}=80~$nV$/\sqrt{\rm Hz}$, and $N_{\rm RIN}(\nu_m)\approx3\times10^{-7}/\sqrt{\rm Hz}$. The figure shows that the expected noise has a minimum for a modulation amplitude $\eta_0\approx10^{-2}$, which is the value normally used.


\section{Experimental setup}
\label{ExpSect}

\begin{figure}[htb]
\begin{center}
\includegraphics[width=11cm]{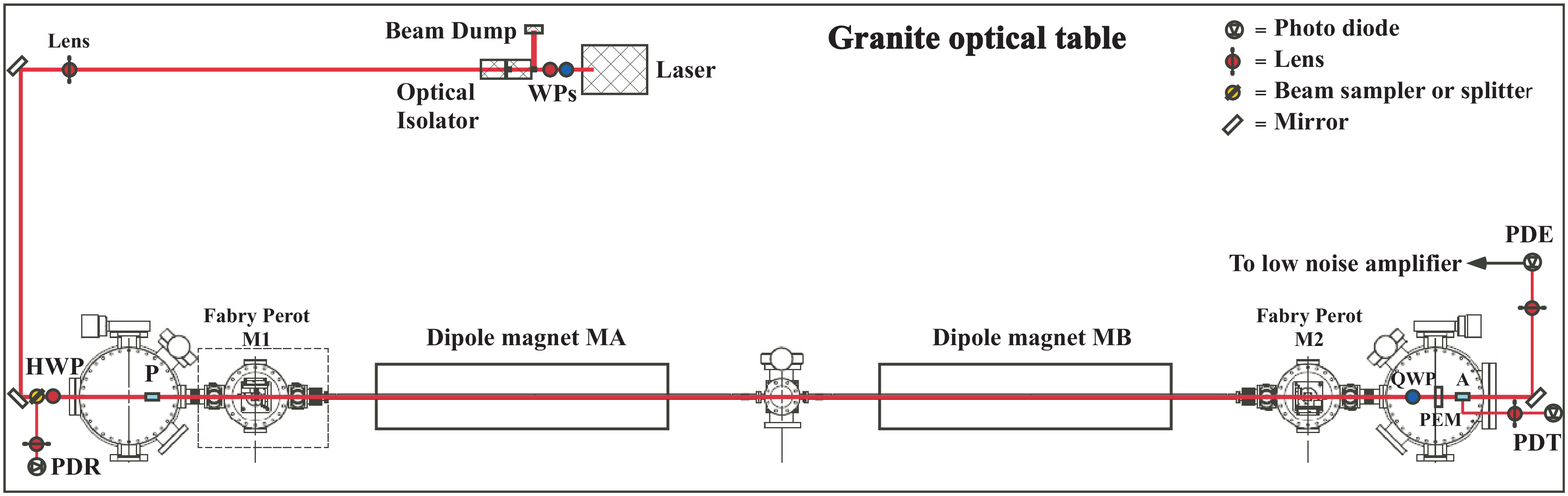}
\includegraphics[width=11cm]{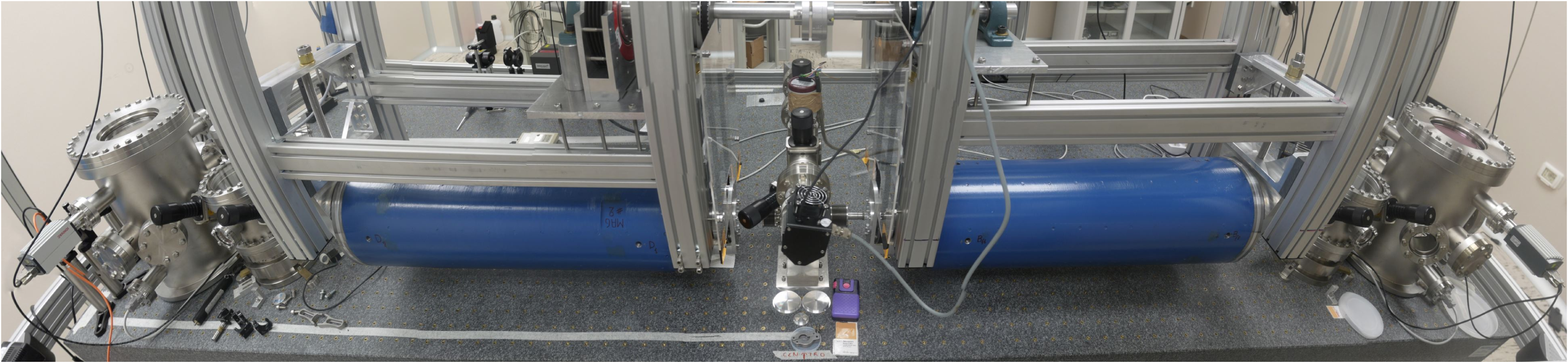}
\end{center}
\caption{Upper panel: optical and mechanical scheme of the apparatus. WPs~=~Wave-plates; HWP~=~Half-wave-plate; PDR~=~Reflection photodiode; P~=~Polariser; Ms~=~Mirrors; QWP~=~Quarter-wave-plate; PEM~=~Photoelastic modulator; A~=~Analyser; PDT~=~Transmission photodiode; PDE~=~Extinction photodiode. Lower panel: A wide-angle picture of the PVLAS apparatus. The two blue cylinders are the permanent magnets.}
\label{Apparatus}
\end{figure}

The upper and lower panels of Figure \ref{Apparatus} show a schematic top view and a photograph of the apparatus. The experiment is hosted inside a class 10,000 clean room. All the optics lay upon a single 4.5~t, $4.8\times1.5\times0.5$~m$^3$ granite honeycomb table. The optical table is seismically isolated from the ground by means of actively operated pneumatic supports. All the mechanical components of the apparatus are made of nonmagnetic materials.

The light source is a 2~W Non Planar Ring Oscillator Nd:YAG laser ($\lambda=1064$~nm), having tuneable emission frequency.
The tuning capabilities of the laser are used to lock the emission frequency of the laser to the resonance frequency of the cavity.
Laser light is mode matched to the Fabry-Perot cavity with a single lens and is linearly polarised immediately before the first mirror. The cavity length is $d=3.303$~m, corresponding to a free spectral range $\nu_{\rm FSR}=45.4$~MHz. The dielectric mirrors, 6~mm thick, 25.4~mm diameter, have fused silica substrates with a radius of curvature of 2~m, and are mounted on $(\theta_x\theta_y\theta_z)$ mirror mounts. The Gaussian cavity mode is TEM00, with a beam radius on the mirrors $w_m=1.2$~mm. The decay time of the cavity has been measured to be $\tau=(2.45\pm0.05)$~ms, corresponding to a finesse of ${\cal F}=\pi c\tau/d\approx700,000$, hence to a path amplification factor $N=445,000$, and to a reflection coefficient $R=0.9999955$. The frequency width of the resonance is 65~Hz, corresponding to a phase interval of less than $10^{-5}$~rad.

The laser frequency is matched to the resonance frequency of the cavity by means of a modified Pound-Drever-Hall feedback system \cite{Cantatore1995}. 
The electronic feedback circuit has the unique feature of allowing the adjustment of the reference point of the loop, equivalent to varying $\delta$ in Equations (\ref{EllipticityAiry}) and (\ref{RotationAiry}). This allows the scanning of the Airy curve of the intensity transmitted by the cavity around its maximum. The amplitude of this interval is in principle limited to the linear range of the error function, but is in practice slightly less. The feedback circuit parameters are controlled by a microprocessor that, in the case the feedback unlocks, re-locks automatically. In a measurement run lasting several days this normally results in a dead time of less then 5\%.

After the cavity, the light crosses the photoelastic ellipticity modulator PEM, that adds a small ellipticity variable at frequency $\nu_m$. In the case of rotation measurements, the quarter-wave-plate QWP is inserted. Finally, the light leaves the polarimeter through the analyser A, that separates the two polarisations. The two beams are collected by the two 1~mm$^2$ InGaAs photodiodes PDT and PDE. The photocurrents are amplified by two low noise transimpedance amplifiers. The extinguished signal is demodulated by two lock-in amplifiers, at $\nu_m$ and at the second harmonic $2\nu_m$. All the relevant signals are properly filtered, digitised, and stored for data analysis.

The magnetic field region is provided by two 96~cm long, 28~cm diameter dipole magnets in Halbach configuration, placed between the mirrors and having a central bore of 20~mm. Each magnet weighs 450~kg. The magnets are sustained by an aluminium structure mechanically decoupled from the rest of the optical table. Overall, the magnets provide a $\int B^2\,d\ell=(10.25\pm0.06)$~T$^2$m. As for the effective length $L$ of each magnet and the value of the magnetic field $B_{\rm ext}$, in the following we will use the FWHM of the function $B^2(z)$, $L=0.82$~m and hence $B_{\rm ext}=2.50$~T. Thus defined, the two magnetic regions are separated by $\approx0.68$~m. The field profile has been shown elsewhere \cite{DellaValle2014CPL}. Stray field on the axis at a position 20~cm outside the magnets is less than 1~G. The magnets can rotate around their axes, at a frequency up to 10~Hz, so that the magnetic field vectors of the two magnets rotate in planes normal to the path of the light stored in the cavity. Two magnetometers, measuring the small stray field of the two magnets, monitor the magnetic field directions.

The synchronous motors driving the two magnets are controlled by two phase-locked signal generators. The same signal generators trigger the data acquisition. The two magnets can rotate at the same frequency with the two magnetic fields making an arbitrary angle, but normally each magnet rotates at its own frequency. In this way the results of one magnet are a countercheck for the results of the other. The two frequencies $\nu_{B1}$ and $\nu_{B2}$ are chosen so to have a common subharmonic whose frequency is used to start data acquisition: at the beginning of each acquisition run, the two magnets have the fields in the same direction. The sampling rate is normally 16 samples/turn for the faster magnet. The rotation frequency of the other magnet is then chosen in such a way that its number of samples/turn contains only factors 2 and 5. A practical example: $\nu_{B1}=8$~Hz, sampling rate $8\times16=128$~Hz, $\nu_{B2}=6.4$~Hz, acquisition start trigger $1.6$~Hz; samples/turn for the second magnet is 20. We have verified that the phase relations between all the generators and the magnets rotation never change during data acquisition.

Two analyses are performed in parallel on the demodulated signal coming from PDE. An online analysis is performed by means of an FFT spectrum analyser. Normally, an integration time of 32~s is chosen and vector averaging is performed between subsequent spectra.
The start trigger ensures that the phases of all the partial spectra are referred to the same angular position of the magnets. This analysis produces visual results in real time, but is not fully exploiting one of the main advantages of the experimental method, namely the frequency selection. In the offline analysis, since all the phases are under control, data acquired in separate time blocks, but with the same experimental conditions, are joined in a single long time series called run. As the time base lengthens, the frequency resolution of the Fourier transform becomes better and better. When doing this, one has to ensure that the $\nu_B$ component of the Fourier transform of the signal from the magnetometer occupies a single frequency bin. This was verified to be true even for the longest runs, having bin size $\Delta\nu\approx1~\mu$s. Time intervals containing anomalous features are expunged from the data. The results of runs differing in the rotation frequency of the magnets or for any other relevant experimental parameter are averaged by using a weighted vector average procedure.

The polarimeter, from the entrance polariser to the analyser, is housed inside a high-vacuum enclosure consisting of five chambers aligned along the light beam path and connected by metallic bellows and by two glass tubes with 12~mm inner diameter traversing the bores of the two magnets. The entrance chamber hosts the polariser P, whereas the exit chamber contains the quarter-wave-plate QWP, the photoelastic modulator PEM, and the analyser A. Each mirror is placed inside a separate chamber, preceded and followed by 10~mm diameter iris diaphragms carved from strongly absorbing glass. The light enters and exits the vacuum through two AR-coated optical glass windows. A system of baffles is placed inside the glass tubes. The central vacuum chamber serves as a pumping station and also contains a central 5~mm diameter diaphragm.

The vacuum system is pumped by turbo-molecular and non-evaporable getter (NEG) pumps, and has a base pressure of less than $10^{-7}$~mbar; the residual atmosphere, monitored by two Residual Gas Analysers, is mainly composed of water vapour, hydrogen and a small amount of methane produced by the NEG pumps. This guarantees that no magnetic birefringence signal from Cotton-Mouton effect on residual gases in the vacuum chamber can interfere with the vacuum measurements \cite{Bakalov1998}. To reduce mechanical vibrations, during measurements in vacuum only the turbo pump of the central chamber is kept on to pump noble gases and methane produced by the NEG pumps. The system can be filled with high purity gases through a leak valve; in this case, the gas pressure is measured with a capacitive transducer. To ensure gas purity, the all-metal gas line is pumped by a turbo pump before gas filling. When the chamber is dosed with noble gases, the NEG pumps are not shut off.

\subsection{Calibration}

The apparatus is calibrated measuring the magnetic linear birefringence of gases (Cotton-Mouton or Voigt effect) \cite{Rizzo1997}. This effect is perfectly analogous to the vacuum magnetic birefringence described by Equation (\ref{birifQED}), but is far more intense already at low gas pressures. The birefringence generated in an atmosphere of gas at pressure $P$ by a magnetic field $B$ is given by the expression \[\Delta n=n_\parallel-n_\perp=\Delta n_u\,\left(\frac{B}{\rm 1~T}\right)^2\,\frac{P}{\rm1~atm},\] where $\Delta n_u$ is the unit birefringence generated in 1~atm of gas by a unitary field $B=1$~T. Typical values of $\Delta n_u$ range from a minimum of $\approx2\times10^{-16}$~T$^{-2}$~atm$^{-1}$ for He \cite{Bregant2009} to $\approx-2.3\times10^{-12}$~T$^{-2}$~atm$^{-1}$ for O$_2$ \cite{Brandi1998} and to $\approx10^{-11}$~T$^{-2}$~atm$^{-1}$ for a few other simple molecules \cite{Rizzo1997}. These measurements give two calibration parameters: the amplitude and the phase of the ellipticity signal. The amplitude can be compared to theoretical calculations as well as to other experimental results, and calibrates the linear response of the polarimeter; the second parameter is the phase of the ellipticity signal, which is determined by the geometry and the electronic response of the apparatus (see Figure \ref{frame}). As seen with the discussion of Figure \ref{frame}, the phase of the signal directly depends on the angle $\theta_0$ of the polariser; this parameter has not a single value during the experiment, but is adjusted from time to time. Electronic components (lock-ins, filters, etc) introduce a phase which depends on the frequency of the signal. The phase of the Cotton-Mouton signals defines what we call the physical phase of the measurements; we expect that the vacuum magnetic birefringence comes with the same phase as the Cotton-Mouton measurement of the noble gases \cite{Rizzo1997}. Any signal in quadrature with respect to the physical phase has to be considered as spurious. As a general principle, all the measured signals are projected onto the physical axis. We explicitly note that the gas measurements are interpreted in terms of a pure birefringence ($\theta=0$). In fact, for gases, no dichroism is associated to a transverse magnetic field; however, a Faraday rotation, due to an eventual small longitudinal component of the field, comes at the magnet rotation frequency $\nu_B$ and not at $2\nu_B$ \cite{Iacopini1983}.

\begin{figure}[htb]
\begin{center}
\includegraphics[width=8cm]{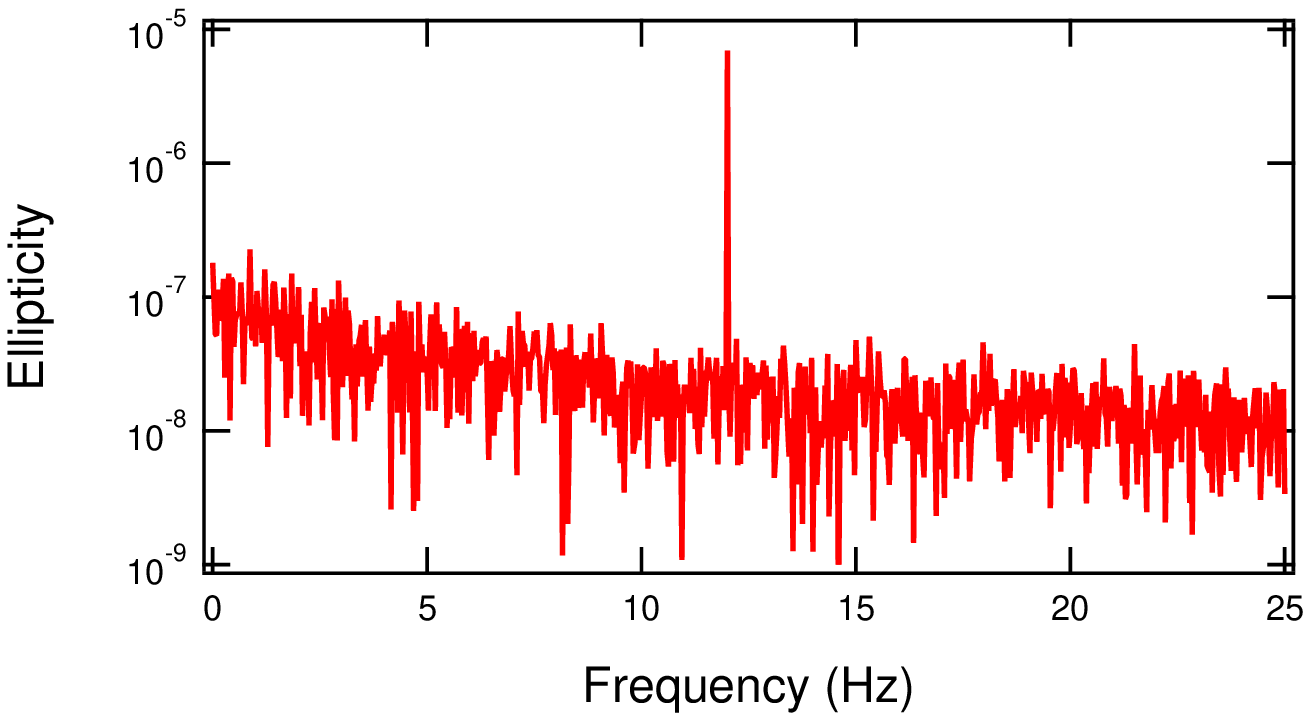}
\includegraphics[width=8cm]{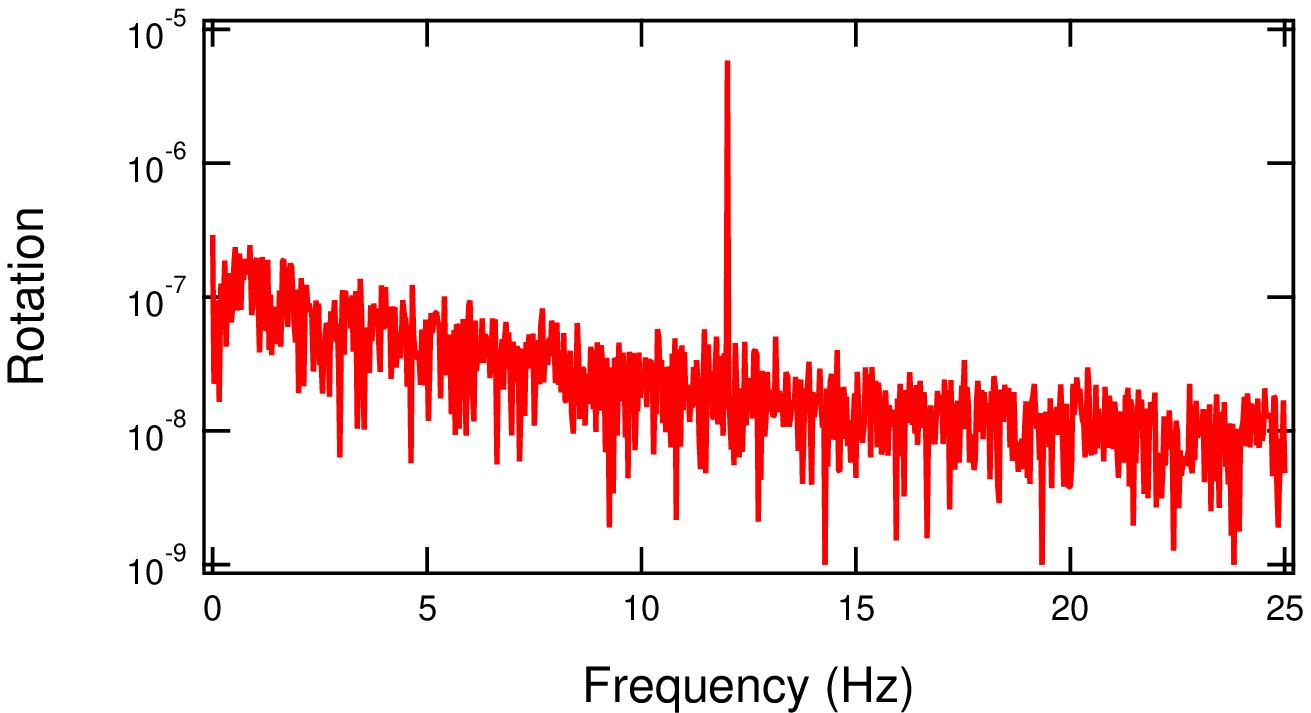}
\end{center}
\caption{Cotton-Mouton effect measurements for 230~$\mu$bar of Ar gas: Fourier spectra of the extinguished intensity demodulated at the modulator frequency $\nu_m$. A single magnet was rotating at $\nu_B=6$~Hz, interesting signals are at $2\nu_B$. Upper panel: ellipticity measurement. Lower panel: rotation measurement. Integration time is $T=640$~s for both spectra.}
\label{CM_spectra}
\end{figure}

In Figure \ref{CM_spectra} we show the spectra of the residual intensity after the analyser, demodulated at the frequency $\nu_m$, with the vacuum chamber filled with 230~$\mu$bar of Ar gas. In the top panel, the Cotton-Mouton ellipticity signal is observed. The bottom panel shows the rotation signal. This indicates that the Fabry-Perot resonances of the two orthogonal polarisation are separated, and the calculations of Section \ref{MirrorBirefringence} apply. Taking the ratio of the amplitudes of the two peaks [see Equations (\ref{EllipticityMeasurement}) and (\ref{RotationMeasurement})] one finds a value $\alpha=3.7~\mu$rad, corresponding to an attenuation factor $k(\alpha)=0.59$. The frequency distance of the two Airy curves is 27~Hz. From these data one can extract a value for the unitary birefringence of Ar gas at room temperature: $\Delta n_u^{\rm (Ar)}=(7.5\pm0.5)\times10^{-15}$~T$^{-2}$~atm$^{-1}$.


\subsection{Studies of the mirrors's wave-plates}
\label{MirrorStudies}

In Section \ref{MirrorBirefringence} we assumed that the axes of the birefringent wave-plates of the two mirrors were always aligned to the input polarisation. 
Here we use a full description of the wave-plates of the two mirrors, placing the second one at an azimuthal angle $\phi_{\rm WP}$ with respect to the first one. We recall \cite{Brandi1997} that the effect of two birefringent wave-plates is equivalent to that of a single wave-plate with a phase difference $\alpha_{\rm EQ}$ given by
\begin{equation}
\alpha_{\rm EQ}=\sqrt{(\alpha_1-\alpha_2)^2 +4\alpha_1\alpha_2\cos^2\phi_{\rm WP}}
\label{alfaEQ}
\end{equation}
and placed at the angle $\phi_{\rm EQ}$ with respect to the slow axis of the first mirror, where
\begin{equation}
\cos2\phi_{\rm EQ}=\frac{\alpha_1/\alpha_2+\cos2\phi_{\rm WP}}{\sqrt{(\alpha_1/\alpha_2-1)^2+4(\alpha_1/\alpha_2)\cos^2\phi_{\rm WP}}}.
\label{phiEQ}
\end{equation}
As noted before, the ratio $R_{\theta,\psi}$ of Equation (\ref{R}) of the ÒspuriousÓ rotation to the ÒtrueÓ ellipticity in Equations (\ref{RotationMeasurement}) and (\ref{EllipticityMeasurement}) is exactly the phase difference (amplified by $-N/2$) of the equivalent wave-plate experienced by the light beam. By varying two of the three quantities: the direction of the mirror axes and the input polarisation direction, one is able to change the phase difference of the equivalent wave-plate of the mirrors while keeping the polarimeter at extinction, namely with the input polarisation aligned with the axis of the equivalent wave-plate. As this procedure changes the equivalent wave-plate, it also changes the ratio of rotation to ellipticity. One is then able to align the fast axis of one mirror wave-plate to the slow axis of the other. In this configuration, if $\alpha_1$ were equal to $\alpha_2$, the resonance curves of the two polarisation auto-states would appear superimposed in a plot like that of Figure \ref{BirefringentAiry}. If $\alpha_1\neq\alpha_2$, the two resonance curves are as near as possible given the difference $\Delta \alpha=\alpha_2-\alpha_1$.

\begin{figure}[htb]
\begin{center}
\includegraphics[width=10cm]{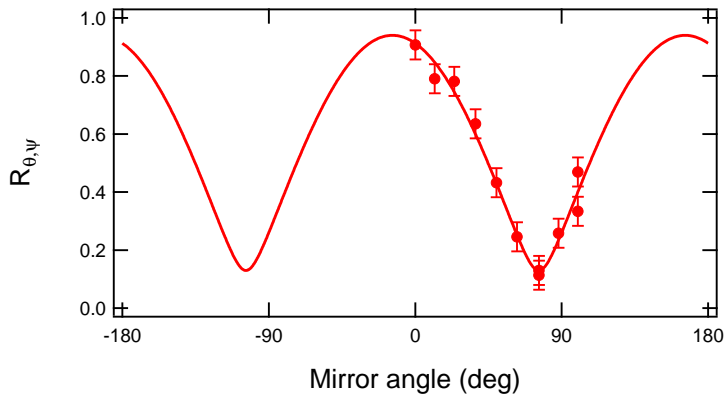}
\end{center}
\caption{Rotation-to-ellipticity signals ratio plotted as a function of the azimuthal angle of the input mirror in a Cotton-Mouton measurement of $230~\mu$bar of Ar gas. The fit line is the ($N/2$-amplified) phase difference $\alpha_{\rm EQ}$ of the equivalent wave-plate of the mirrors given by Equation (\ref{alfaEQ}).}
\label{MirrorWP}
\end{figure}

In Figure \ref{MirrorWP}, we show the ratio of the values of rotation to ellipticity in a Cotton Mouton measurement, plotted as a function of the azimuthal angle of the first mirror. Each rotation step, of about $15^\circ$, has been followed by cavity realignment through the adjustment of the two tilt stages of the mirror, by optimisation and measurement of the extinction ratio, and by measurement of the finesse. The experimental points are fitted with Equation (\ref{R}), where $\alpha$ is given by Equation (\ref{alfaEQ}). The best fit produces values for the quantities $N\alpha_1/2$, $N\alpha_2/2$, and for the angular position of the maxima with respect to the initial angular position of the input mirror ($\phi_{\rm WP}=0$). With $N/2\approx2.2\times10^5$, the phase differences of the two mirrors are calculated to be $(2.4\pm0.1)~\mu$rad and $(1.9\pm0.1)~\mu$rad. From this fit only it is not possible to label each mirror with its phase difference for reflection. According to the relative angular position of the two mirrors, the value of $\alpha_{\rm EQ}$ can be found between $0.6~\mu$rad and $4.3~\mu$rad, which is equivalent to saying that the Airy curve of the ellipticity resonance is 5 to 31~Hz away from the resonance of the input polarisation. 

\begin{figure}[htb]
\begin{center}
\includegraphics[width=10cm]{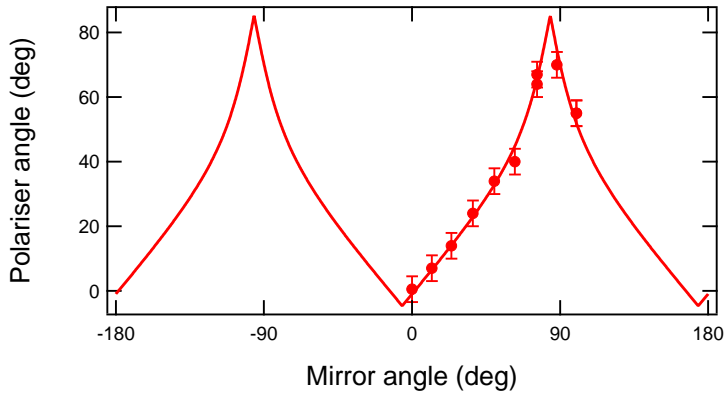}
\end{center}
\caption{Polariser angle as a function of the azimuthal angle of the mirror in a Cotton-Mouton measurement of $230~\mu$bar of Ar. Data are fitted with $\phi_{\rm EQ}$ as given by Equation (\ref{phiEQ}).}
\label{MirrorAxis}
\end{figure}

In Figure \ref{MirrorAxis}, the values taken by the polariser angle while tracking the best extinction ratio in the process described above are plotted against the input mirror angle. The curve is fitted with Equation (\ref{phiEQ}). The best fit produces a value $\alpha_1/\alpha_2=0.62\pm0.08$, allowing the assignment of the phase delay of each mirror. 
This value is slightly different from the one obtained by the fit in Figure \ref{MirrorWP}, but is compatible within the fit uncertainties. However, the zero references of $\phi_{\rm WP}$ in the two fits appear to be different by about $10^\circ$, well beyond the fit uncertainty. This might be due to the presence of other birefringent elements (mirror substrates and PEM) between the two crossed polarisers. As these elements are fixed during the measurement, while the equivalent wave-plate of the mirrors is varying, their contribution to the total anisotropy varies from one measurement to the other. The position of the polariser tracks the position of the equivalent wave-plate of all the wave-plates of the system, and not only of that of the mirrors. On the contrary, the data of Figure \ref{MirrorWP}, being the ratio of signals at $2\nu_B$, do not suffer from the same problem. Anyway, the smallness of the difference of the two determinations of the reference angle indicates that the importance of birefringent elements other than the reflecting surface of the mirrors is very limited.

\begin{figure}[htb]
\begin{center}
\includegraphics[width=10cm]{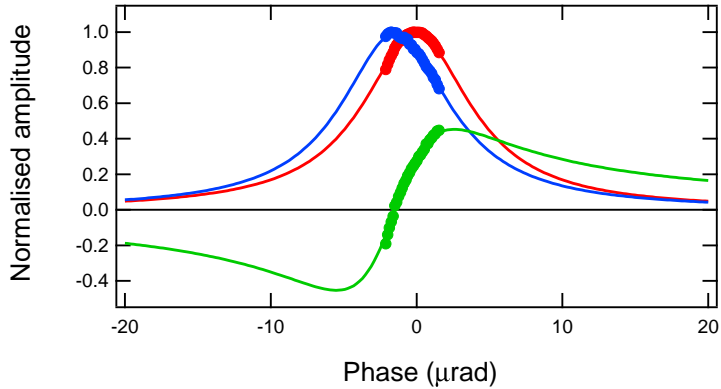}
\end{center}
\caption{From left to right: ellipticity, transmitted intensity, and rotation, measured for $230~\mu$bar of Ar gas, plotted as functions of the set point of the laser locking feedback circuit. The continuous lines are the fits obtained with formulas (\ref{EllipticityMeasurement}), (\ref{transmissionFP}), and (\ref{RotationMeasurement}).}
\label{BirefringentFP}
\end{figure}

A unique feature of our apparatus is the possibility to change the set point of the feedback electronic circuit that locks the laser frequency to the resonance frequency of the cavity. This allows to perform polarimetric measurements with arbitrary values of $\delta$, in this way fully testing the mathematics presented in Section \ref{MirrorBirefringence}. In Figure \ref{BirefringentFP}, we show an experimental realisation of Figure \ref{BirefringentAiry}. The continuous lines are the fits of the data obtained with formulas (\ref{transmissionFP}), (\ref{EllipticityMeasurement}), and (\ref{RotationMeasurement}). In the three fits, a single value of the resonance width has been used. Ellipticity and rotation curves are forced to have the same centre of resonance and the same amplitude coefficient. The fit determines the scale factor between the feedback set point and the phase $\delta$. The distance between the two Airy curves is found to be $\alpha=1.5~\mu$rad (with negative sign), corresponding to a frequency difference of the two resonance frequencies of about 11~Hz.

\section{Vacuum measurements results and discussion}
\label{Misure}

\begin{table}[htb]
\begin{center}
\begin{tabular}{|c|c|c|r@{~}l|c|c|c|}
\hline
Run \# & Quantity & Magnets & \multicolumn{2}{c|}{$2\nu_B$} & $T$ (s)         & ${\cal F}$      & $k(\alpha)$ \\
\hline
0  & $\psi$   & MA+MB   &   &                           & $6.7\times10^5$ & $6.7\times10^5$ & 0.50        \\
1  & $\psi$   & MB      &  8&Hz                         & $1.0\times10^6$ & $7.0\times10^5$ & 0.65        \\
2  & $\psi$   & MA      & 10&Hz                         & $1.0\times10^6$ & $7.0\times10^5$ & 0.65        \\
3  & $\psi$   & MB      & 10&Hz                         & $8.9\times10^5$ & $7.0\times10^5$ & 0.65        \\
4  & $\psi$   & MA      & 12.5&Hz                       & $8.9\times10^5$ & $7.0\times10^5$ & 0.65        \\
\hline
5  & $\theta$ & MA+MB   & 10&Hz                         & $1.4\times10^5$ & $7.0\times10^5$ & 0.65        \\
\hline
\end{tabular}
\end{center}
\caption{Experimental conditions for the runs in vacuum. In the ``\#0 run", taken from Reference \cite{DellaValle2014PRD}, the magnet rotation frequency ranged from 2.4 to 3~Hz. $T$ is the integration time.}
\label{runs}
\end{table}

\begin{figure}[htb]
\begin{center}
\includegraphics[width=5cm]{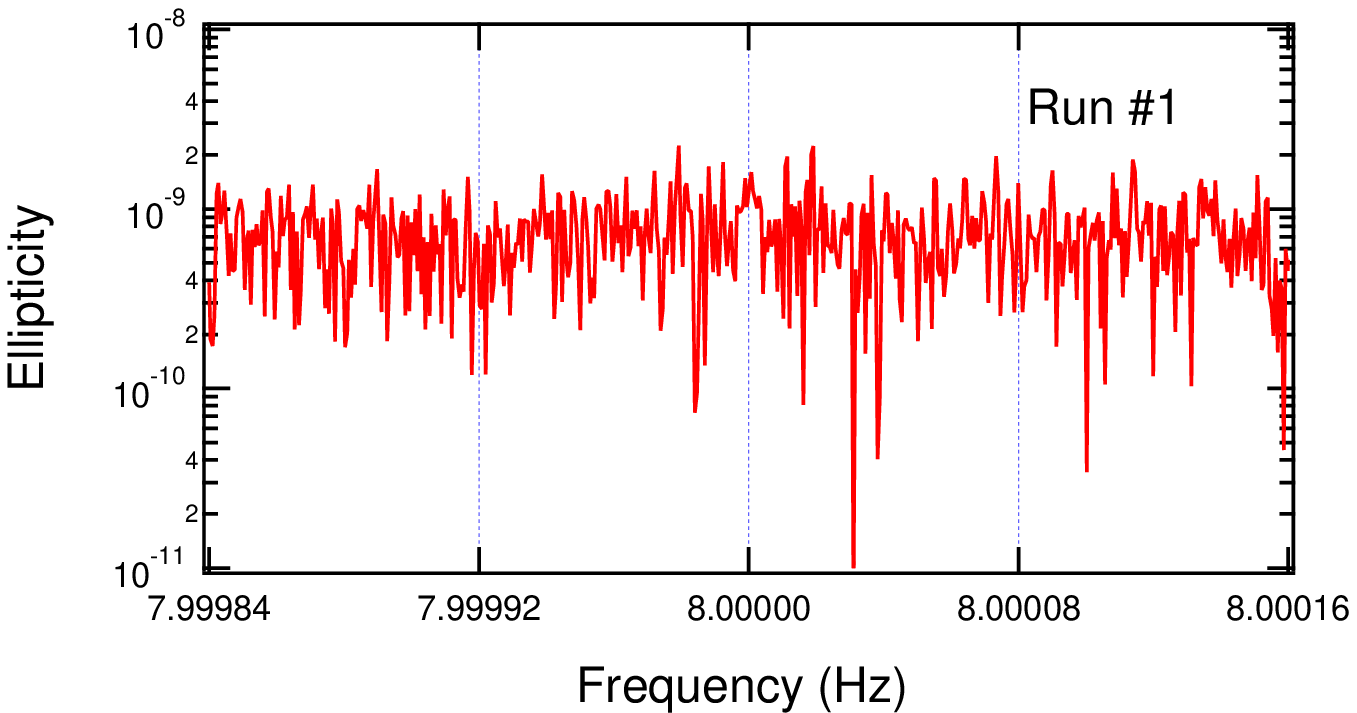}
\includegraphics[width=5cm]{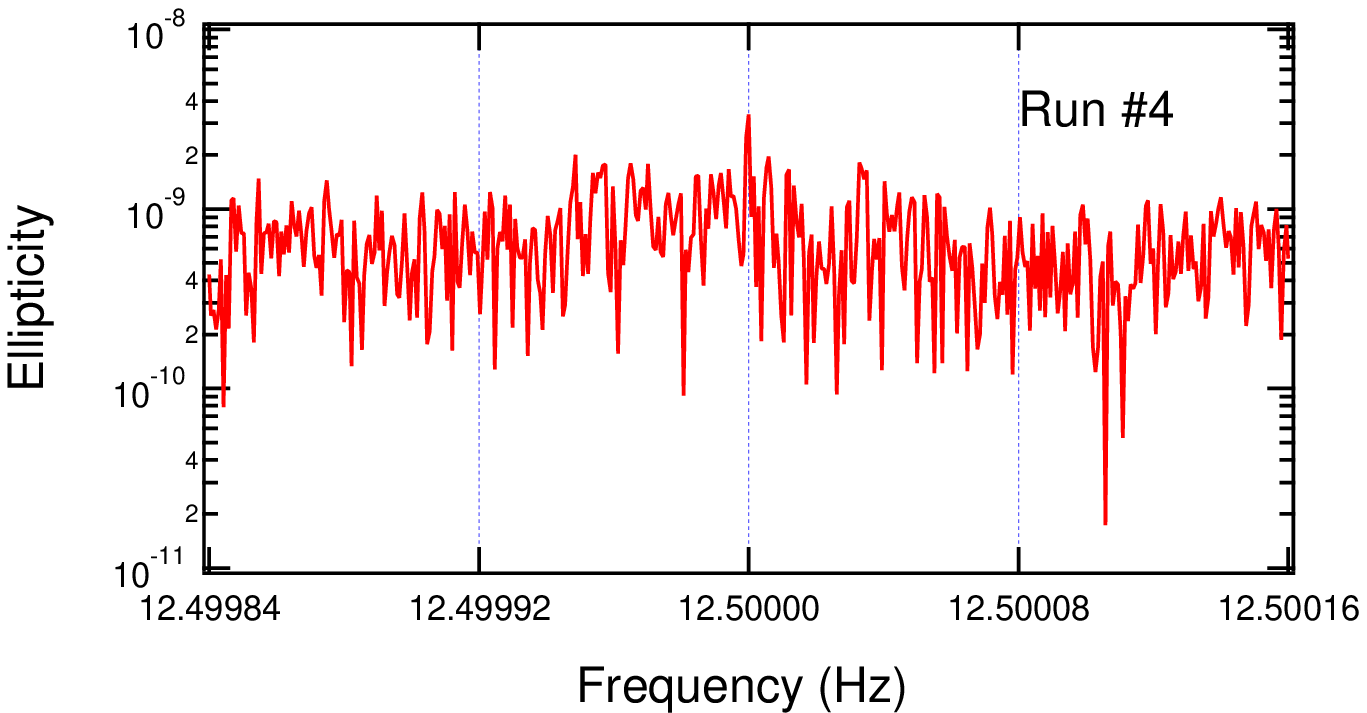}
\end{center}
\caption{Fourier transform of the ellipticity signals of run \#1 and \#4. In both cases, a structure is present around $2\nu_B$. These data do not contribute to the results presented in this work.}
\label{discarded}
\end{figure}

In this Section we present the ellipsometric measurements carried out on vacuum in the attempt to test its opto-magnetic properties. The runs considered in this work are listed in Table \ref{runs}. Since the measurements have been taken making use of a birefringent cavity, the ellipticity data can be interpreted also in terms of rotation; the converse is also true. 
The integrated noise level in the ellipticity measurement allow to cast upper limits on the magnetic birefringence predicted by QED, and also on the existence of hypothetical particles coupling to two photons, ALPs and milli-charged particles. Two ellipticity runs, at $\nu_B=4$~Hz and at $\nu_B=6.25$~Hz with integration time $T=10^6$~s and $T=8.9\times10^5$~s have been discarded due to the presence of spurious structures in the Fourier transform of the signals around $2\nu_B$ (see Figure \ref{discarded}). In fact, a signal coming from a magnetic birefringence cannot occupy more than a single bin. These structures are the consequence of a misalignment of the glass tube. We have developed an alignment procedure that prevents the appearance of systematic peaks in the spectra, but this does not prevent a small drift of its position during the long runs.

\begin{figure}[htb]
\begin{center}
\includegraphics[width=5.8cm]{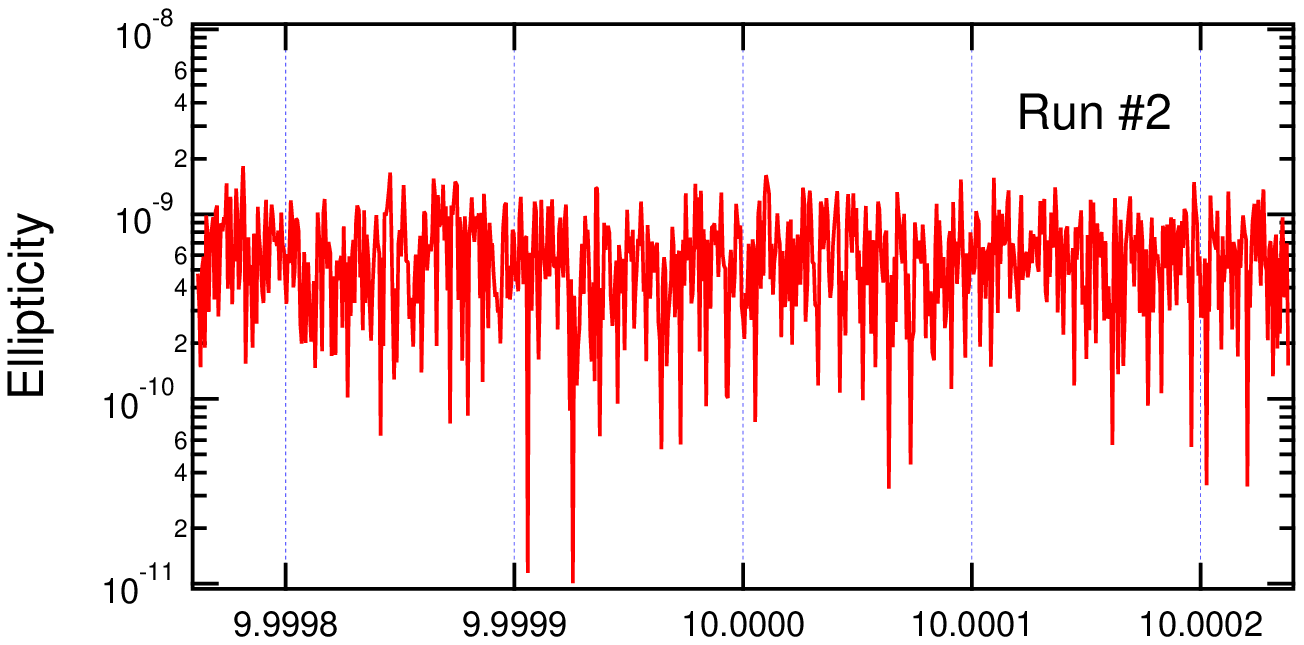}
\includegraphics[width=5.8cm]{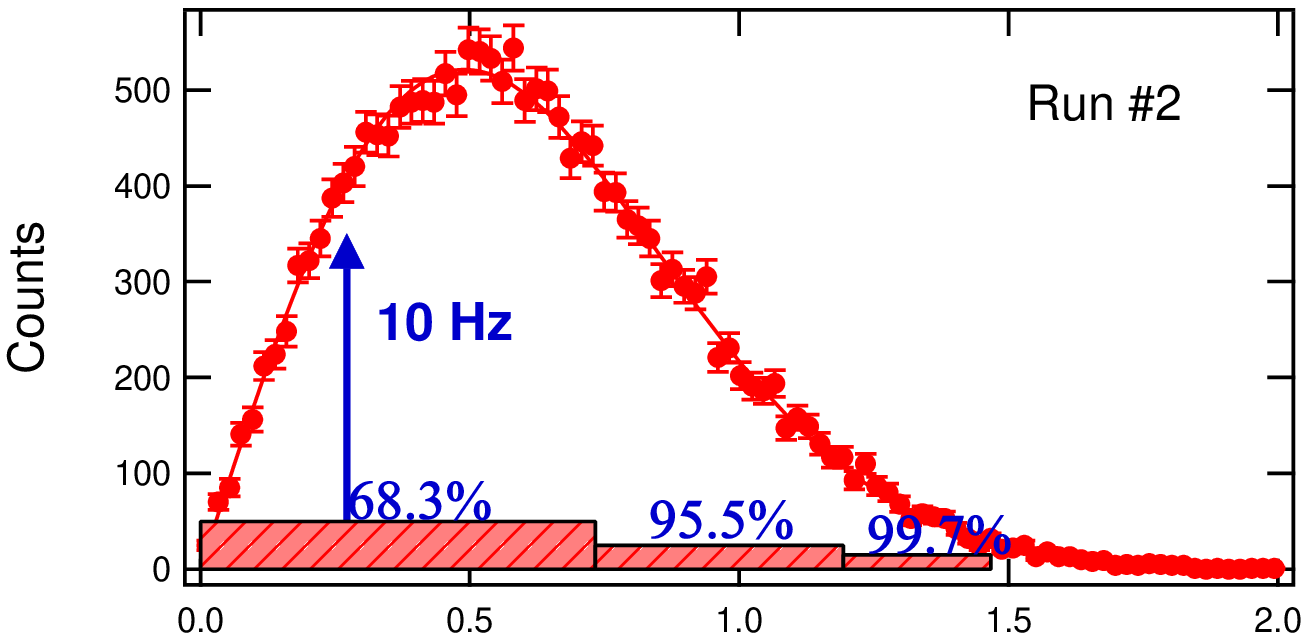}
\includegraphics[width=5.8cm]{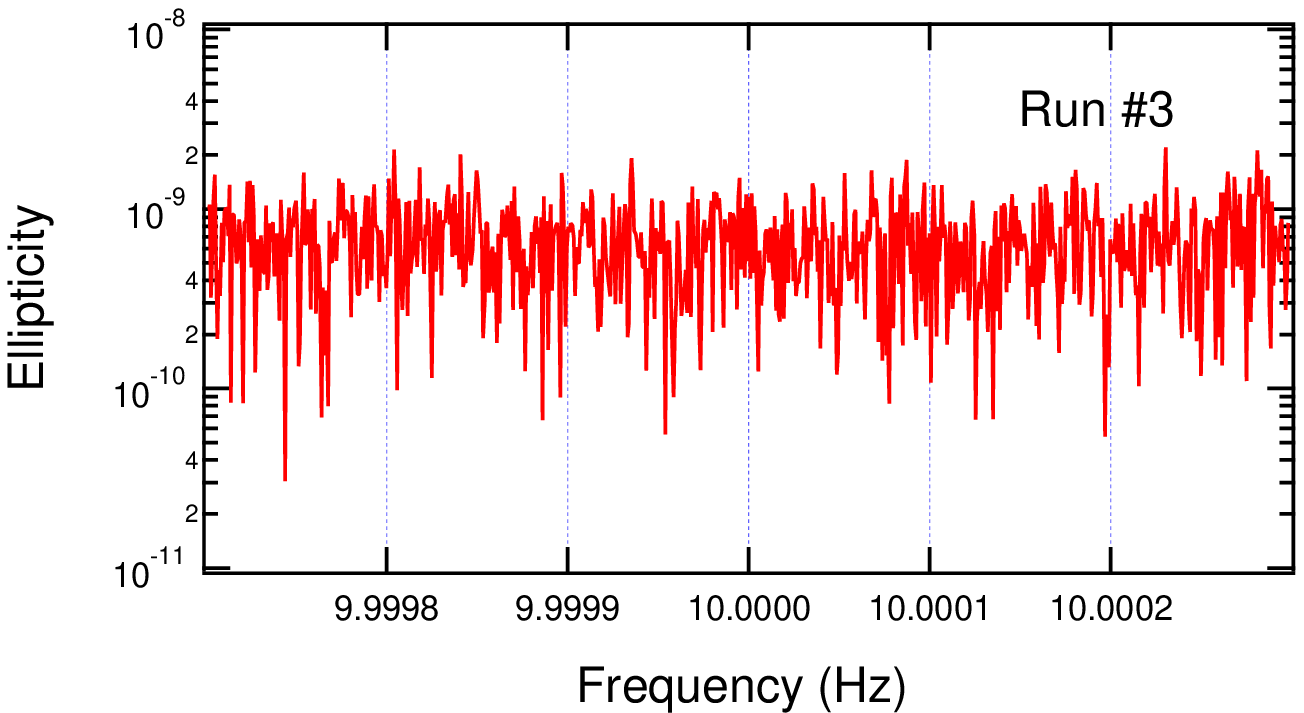}
\includegraphics[width=5.8cm]{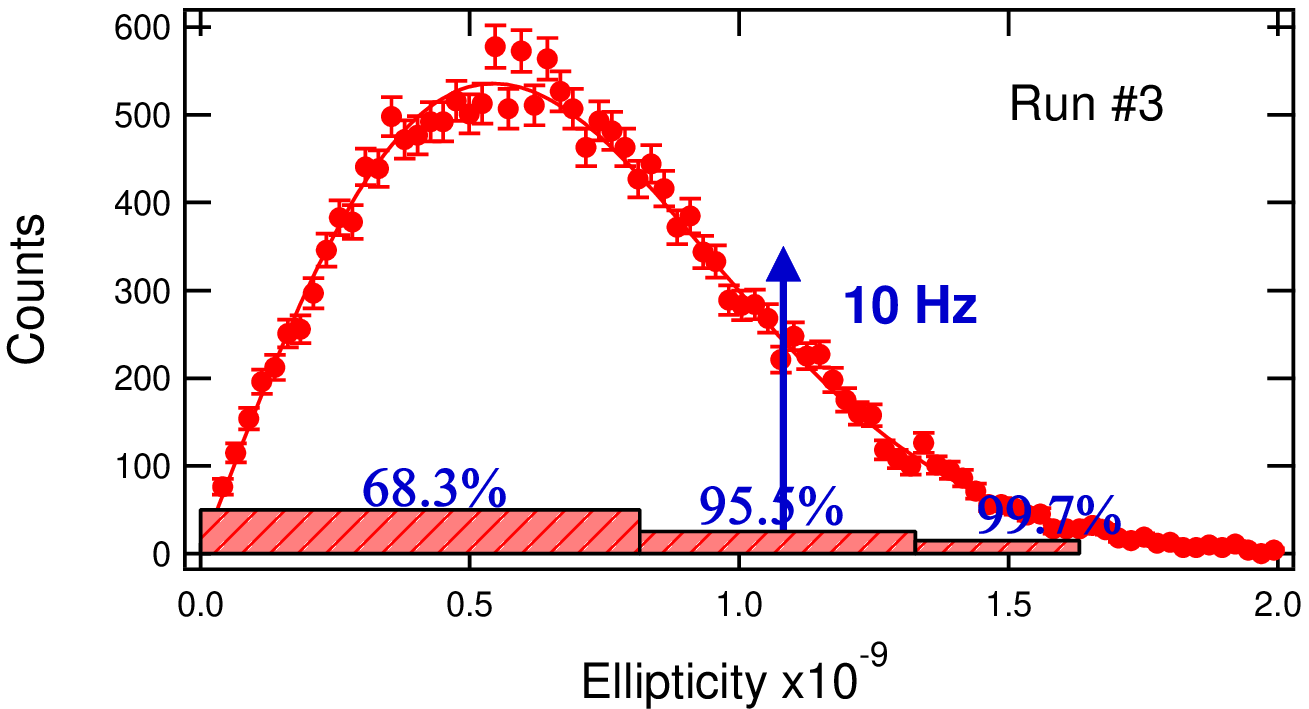}
\vspace{1cm}
\includegraphics[width=5.8cm]{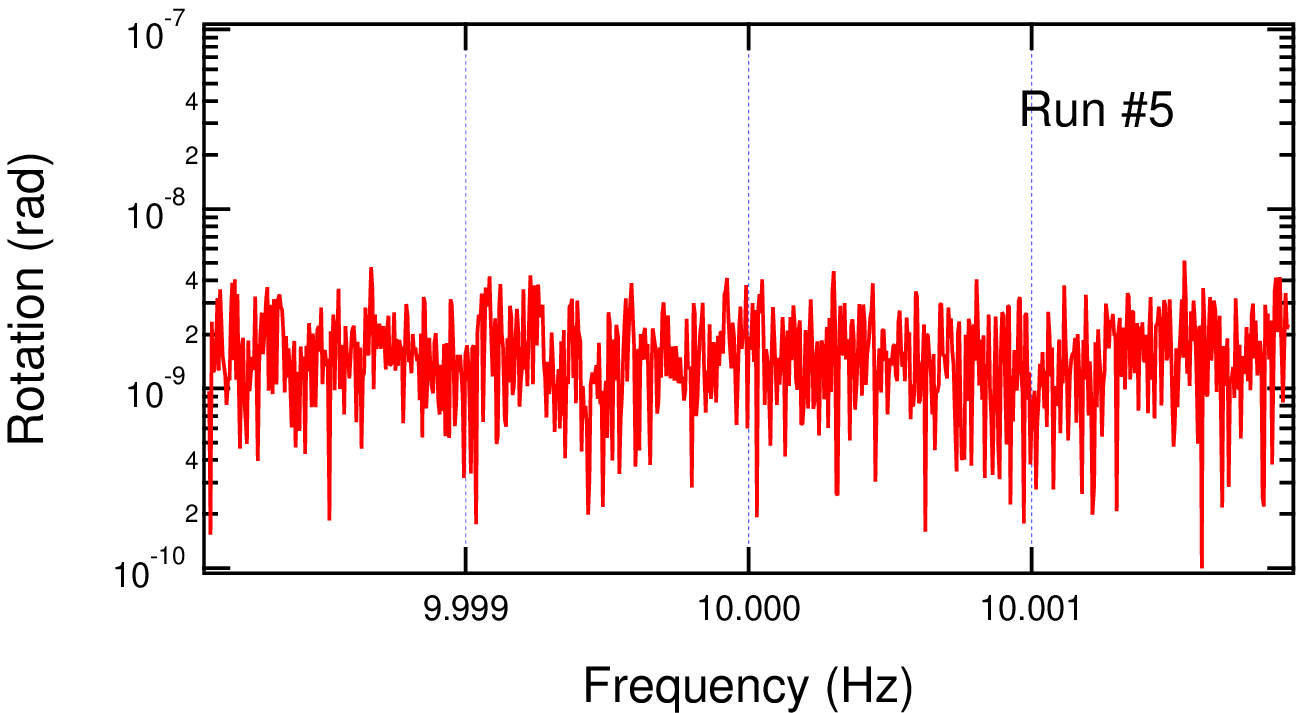}
\includegraphics[width=5.8cm]{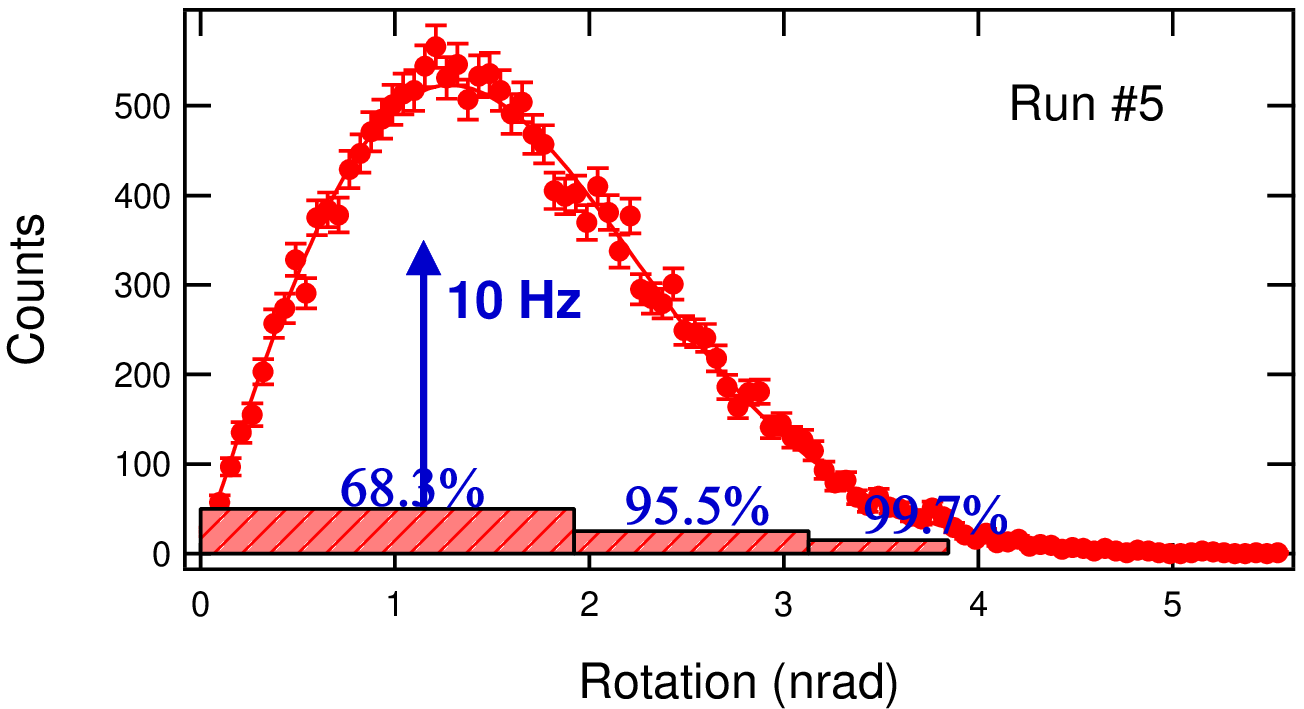}
\end{center}
\caption{Ellipticity and rotation runs in vacuum. First column: amplitude of the complex Fourier transforms of the signal in a narrow interval around $2\nu_B$. The values are corrected for the $k(\alpha)$ factor. Second column: histograms of the values plotted in the first column; the data are fitted with the Rayleigh distribution, the vertical arrows mark the unprojected values at $2\nu_B$. The strips at the bottom of the plots correspond to the 68.3\%, 95.5\%, and 99.7\% integrated probabilities. First row: one magnet rotating at 5~Hz, with integration time $T=10^6$~s. Second row: one magnet rotating at 5~Hz, $T=8.9\times10^5$~s. Third row: two magnets rotating at 5~Hz.}
\label{VacuumEllipticity}
\end{figure}


In Figure \ref{VacuumEllipticity} the results of all the runs are shown. In the left column of plots, the amplitudes of the complex Fourier transform of the signal in a narrow interval around $2\nu_B$ show the absence of any structure due to spurious signals. The values at $2\nu_B$, projected along the physical axis, represent the results of the measurement. In the right column of plots, the histograms of the ellipticity noise amplitude values plotted on the left are shown, fitted with the Rayleigh distribution \[P_{\rm R}(\rho)=\frac{\rho}{\sigma^2}\,e^{-\frac{\rho^2}{2\sigma^2}}\] of a two-dimensional variable $\rho=\sqrt{x^2+y^2}$, where $x$ and $y$ are two independent Gaussian variables having the same standard deviation $\sigma$. In our case, $x$ and $y$ are the projection of the complex Fourier components of the signal onto the physical and the quadrature axes. The values obtained for $\sigma$ define the noise level of the measurement for an integration time $T$. The measured sensitivity $S_{2\nu_B}^{\rm meas}$ of the apparatus at the frequency of interest is then \[S_{2\nu_B}^{\rm meas}=\sqrt{T}\,\sigma.\]


\begin{table}[htb]
\begin{center}
\begin{tabular}{|c|c|r@{$\times$}l|r@{}l|r@{$\times$}l|c|}
\hline
Run \# & Quantity       & \multicolumn{2}{c|}{In-phase} & \multicolumn{2}{c|}{Quadrature} & \multicolumn{2}{c|}{$\sigma$} & $S_{2\nu_B}^{\rm meas}$ $\left(1/\sqrt{\rm Hz}\right)$ \\
\hline
0  & $\psi$         & $+5.2$&$10^{-10}$             & $+6.5\times$&$10^{-10}$            & $2.6$&$10^{-9}$               & $2.1\times10^{-6}$ \\
2  & $\psi$         & $-6.9$&$10^{-11}$             & $+2.6\times$&$10^{-10}$            & $4.9$&$10^{-10}$              & $4.9\times10^{-7}$ \\
3  & $\psi$         & $-4.1$&$10^{-10}$             & $+1.0\times$&$10^{-9}$             & $5.4$&$10^{-10}$              & $5.1\times10^{-7}$ \\
5  & $\theta$ (rad) & $-6.6$&$10^{-11}$             & $-1.9\times$&$10^{-9}$             & $1.3$&$10^{-9}$               & $4.8\times10^{-7}$ \\
\hline
0' & $\theta$ (rad) & $+5.2$&$10^{-10}$             &&                                   & $2.6$&$10^{-9}$               & $2.1\times10^{-6}$ \\
2' & $\theta$ (rad) & $-9.4$&$10^{-11}$             &&                                   & $6.7$&$10^{-10}$              & $6.7\times10^{-7}$ \\
3' & $\theta$ (rad) & $-5.6$&$10^{-10}$             &&                                   & $7.4$&$10^{-10}$              & $6.9\times10^{-7}$ \\
5' & $\psi$         & $+9.0$&$10^{-11}$             &&                                   & $1.8$&$10^{-9}$               & $6.5\times10^{-7}$ \\
\hline
\end{tabular}
\end{center}
\caption{Ellipticity and rotation results for all the runs in vacuum. The first five lines refer to the measurements actually performed. The lower half of the table, with primed run numbers, reports the values obtained through the use of Equations (\ref{EllipticityMeasurement}) and (\ref{RotationMeasurement}).}
\label{Results}
\end{table}

\begin{table}[htb]
\begin{center}
\begin{tabular}{|c|c|r@{$\times$}l|r@{}l|r@{$\times$}l|c|}
\hline
Run \# & Quantity       & \multicolumn{2}{c|}{In-phase} & \multicolumn{2}{c|}{Quadrature} & \multicolumn{2}{c|}{$\sigma$} & $S_{2\nu_B}^{\rm meas}$ $\left(1/\sqrt{\rm Hz}\right)$ \\
\hline
0  & $\Delta n$     & $+2.5$&$10^{-22}$             & $+3.1\times$&$10^{-22}$            & $1.3$&$10^{-21}$              & $1.0\times10^{-18}$ \\
2  & $\Delta n$     & $-6.4$&$10^{-23}$             & $+2.4\times$&$10^{-22}$            & $4.5$&$10^{-22}$              & $4.5\times10^{-19}$ \\
3  & $\Delta n$     & $-3.8$&$10^{-22}$             & $+9.3\times$&$10^{-22}$            & $5.0$&$10^{-22}$              & $4.7\times10^{-19}$ \\
5' & $\Delta n$     & $+4.2$&$10^{-23}$             &&                                   & $8.2$&$10^{-22}$              & $3.0\times10^{-19}$ \\
\hline
0' & $\Delta\kappa$ & $+2.5$&$10^{-22}$             &&                                   & $1.3$&$10^{-21}$              & $1.0\times10^{-18}$ \\
2' & $\Delta\kappa$ & $-8.7$&$10^{-23}$             &&                                   & $6.2$&$10^{-22}$              & $6.2\times10^{-19}$ \\
3' & $\Delta\kappa$ & $-5.2$&$10^{-22}$             &&                                   & $6.8$&$10^{-22}$              & $6.4\times10^{-19}$ \\
5  & $\Delta\kappa$ & $-3.1$&$10^{-23}$             &$-8.8\times$&$10^{-22}$             & $6.0$&$10^{-22}$              & $2.2\times10^{-19}$ \\
\hline
\end{tabular}
\end{center}
\caption{Determinations of the magnetic birefringence and dichroism of vacuum for $B=2.5$~T. The primed measurements are obtained through the use of Equations (\ref{EllipticityMeasurement}) and (\ref{RotationMeasurement}).}
\label{BirefringenceDichroism}
\end{table}

In the first half of Table \ref{Results} we summarise the results of all the measurements in vacuum. Due to the mixing of ellipticity and rotation, each line can be interpreted also in terms of the reciprocal quantity. The second half of the same table, with primed run numbers, presents the values obtained by applying Equations (\ref{EllipticityMeasurement}) and (\ref{RotationMeasurement}). The lines marked with $\psi$ give four determinations of the magnetic birefringence of vacuum; as many determinations of the dichroism are given by the lines marked with $\theta$. These numbers are listed in Table \ref{BirefringenceDichroism}.\footnote{One must note that the measured $\psi$ and $\theta$ are intrinsically integral quantities. As a consequence, the values of $\Delta n$ and $\Delta\kappa$ in the table are not point functions, but average quantities. Moreover, they are calculated with the length of the magnets defined for convenience as the FWHM of $B^2(z)$. Hence, they have a precise meaning only in the cases in which their expression is proportional to $B_{\rm ext}^2$, namely the QED vacuum and the birefringence of ALPs and MCPs in the limit of large mass.} The weighted averages of the numbers listed in the ``In-phase" column of Table \ref{BirefringenceDichroism} are
\begin{eqnarray}
\label{BirefringencePVLAS}
&\Delta n^{\rm (PVLAS)}=(-1.5\pm3.0)\times10^{-22}\qquad&@~B=2.5{\rm~T},\\
&\Delta\kappa^{\rm (PVLAS)}=(-1.6\pm3.5)\times10^{-22}\qquad&@~B=2.5{\rm~T}.
\label{DichroismPVLAS}
\end{eqnarray}
 The quadrature value of $\Delta n$ results to be $(+5.2\pm3.2)\times10^{-22}$. All the numbers found are compatible with zero. The value of $\Delta n^{\rm (PVLAS)}$ is an order of magnitude larger than the birefringence predicted by QED [Equation (\ref{DeltanQED})] and serves only as an upper limit.

\begin{figure}[htb]
\begin{center}
\includegraphics[width=10cm]{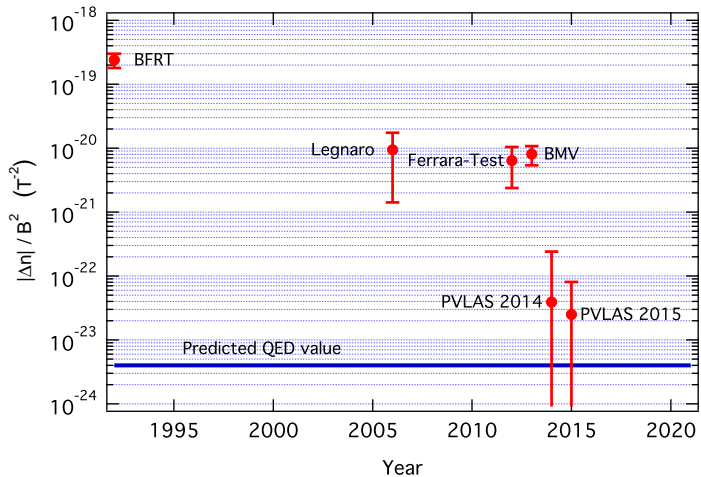}
\end{center}
\caption{Time evolution of the measurement of vacuum magnetic birefringence normalised to $B_{\rm ext}^2$. Error bars correspond to $1\sigma$. Values have been taken from the following references: BFRT \cite{Cameron1993}; Legnaro \cite{Bregant2008}, Ferrara Test \cite{DellaValle2013}, BMV \cite{BMV2014}, PVLAS 2014 \cite{DellaValle2014PRD}.}
\label{TimeEvolution}
\end{figure}

Figure \ref{TimeEvolution} shows the time evolution of the measurement of the QED magnetic birefringence of vacuum. To compare the different experiments, the measured values of the birefringence have been normalised to $B_{\rm ext}^2$. By extrapolation, one could predict that it should not take too long before the measurement is performed successfully. Anyway, this will not happen if the sensitivity of the polarimeter will not improve by an order of magnitude. The next section briefly discusses the noise issue.

\subsection{Noise considerations}

The values found for the sensitivity of the polarimeter (see last column of Table \ref{Results}) are a factor four better than the values obtained in previous versions of the experiment \cite{DellaValle2014PRD}, but is still far from the theoretical value $6\times10^{-9}~1/\sqrt{\rm Hz}$ that is computed by adding all the known noise sources, as in Figure \ref{IntrinsicNoise}.

It is not clear which could be the sources of the excess noise. A few things are known, though: first of all, the noise comes from the cavity; in fact, when the mirrors are removed, the polarimeter performance is limited only by intrinsic noise; this would exclude the laser as a source of noise. Since we are talking of noise in ellipticity and rotation, one must find a mechanism that produces noise in these two quantities.

A possible source of noise is the intrinsic birefringence of the mirrors. One could imagine a few mechanisms for a wide band modulation of this parameter. One of them could be mechanical movement of the mirrors induced by seismic noise: as the surface of such mirrors has a birefringence pattern both in amplitude and in axis direction \cite{Micossi1993}, one could imagine that environmental mechanical noise moves the beam spot on the surface of the mirror, modulating the birefringence in a wide frequency range. However, this mechanism can be excluded: the amplitude of the ellipticity signal generated by forcing the optical bench to oscillate at a single frequency with known amplitude was measured and compared to the observed mechanical noise floor at $2\nu_B$. The measurement was repeated for the three spatial directions; in all cases the observed noise floor was found much too weak to account for the observed sensitivity of the polarimeter. Moreover, no improvement of the sensitivity was observed when the ellipsometer was running in the quietest situations (during nights, with air conditioning switched off, etc.).

Considering again the intrinsic birefringence, another mechanism that could be invoked to explain the sensitivity is the insufficient thermal stability of the mirrors \cite{Chui1995}. This mechanism would imply a dependence of the sensitivity upon the light power inside the cavity. Such a dependence is observed only for frequencies below $\approx1$~Hz. Nonetheless, we are planning to cool the mirrors down to the liquid nitrogen temperature.

A notable aspect of the observed noise, is that it is quite independent from the value of the coefficient $k(\alpha)$, as was observed during the rotation of the mirrors reported in the previous section. This seems to indicate that the noise may originate from diffused light inside the polarimeter and have nothing to do with intrinsic birefringence of the mirrors. However, the system of optical baffles and diaphragms that was installed along the beam path was able to get rid of the spurious signals at frequency $2\nu_B$ that haunted the measurements in the past \cite{DellaValle2013}, but seems not to have benefited the wide band noise. Further studies are ongoing.

\subsection{Limits on hypothetical particles}

\begin{figure}[htb]
\begin{center}
\includegraphics[width=10cm]{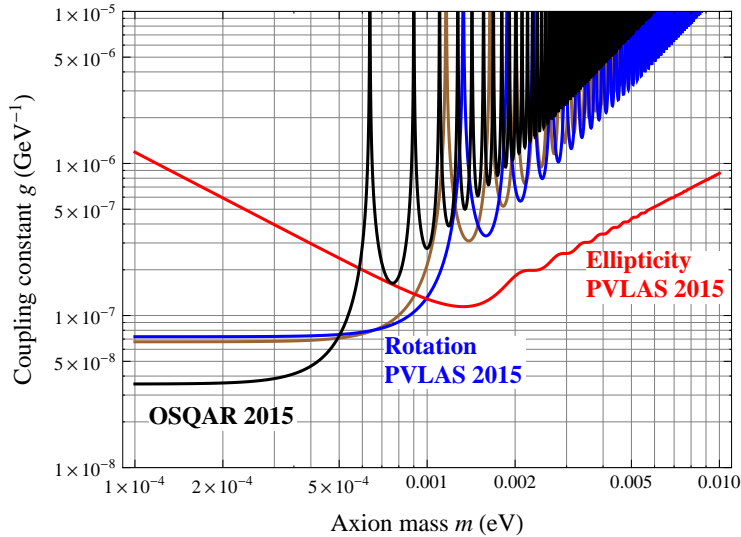}
\end{center}
\caption{Exclusion plot for ALPs particles from laboratory experiments at 95\% c.l. The excluded region is above the curves. The limits hold for both scalar and pseudoscalar ALPs. Besides the PVLAS results, the figure shows also the measurements by OSQAR \cite{OSQAR2015} and ALPS \cite{ALPS2010} collaborations.}
\label{ALPsLimits}
\end{figure}

The measurements of ellipticity and rotation can be used to draw an exclusion plot in the plane $(m,g)$ for Axion-like particles. One must note, however, that it is not possible to average together measurements taken with different magnet lengths [cf. Equations (\ref{ALPs})]. The best limits we can provide derive from the ellipticity measurements taken with one rotating magnet (run \#2 and \#3 in Table \ref{BirefringenceDichroism}) and the rotation measurements taken with two magnets (run \#0' and \#5). The results are shown in Figure \ref{ALPsLimits}. The limits hold for both scalar and pseudoscalar ALPs. Below $0.5$~meV, the most stringent results are given by a recent measurement by the OSQAR experiment \cite{OSQAR2015}, whereas our ellipticity measurement dominates the $m\geq1$~meV region. Between these two values, our rotation measurement almost coincides with the 2010 ALPS result \cite{ALPS2010}. One must obviously remind that the whole region has already been excluded by the CAST solar helioscope down to the level $g\sim10^{-10}$~GeV$^{-1}$ \cite{CAST2007}. The interest for the laboratory experiments resides in the fact that their results are model independent.

\begin{figure}[htb]
\begin{center}
\includegraphics[width=10cm]{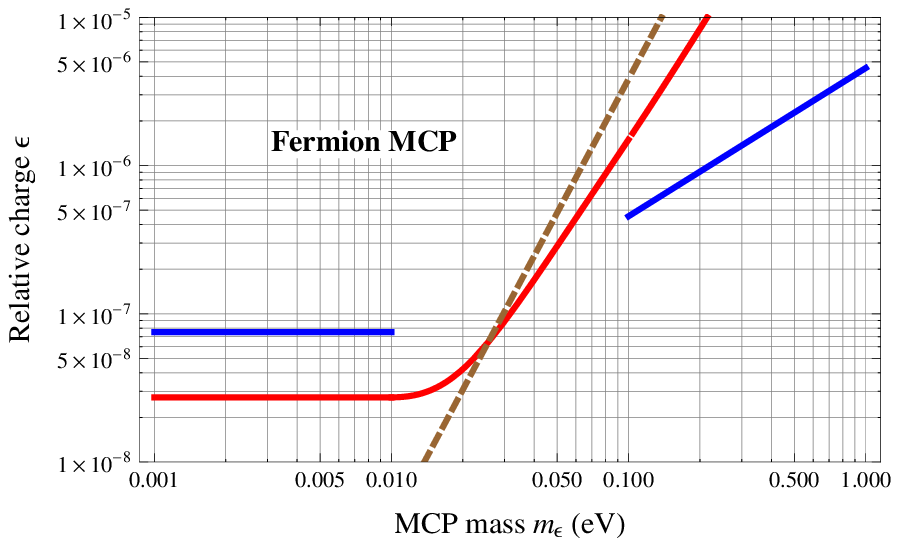}
\includegraphics[width=10cm]{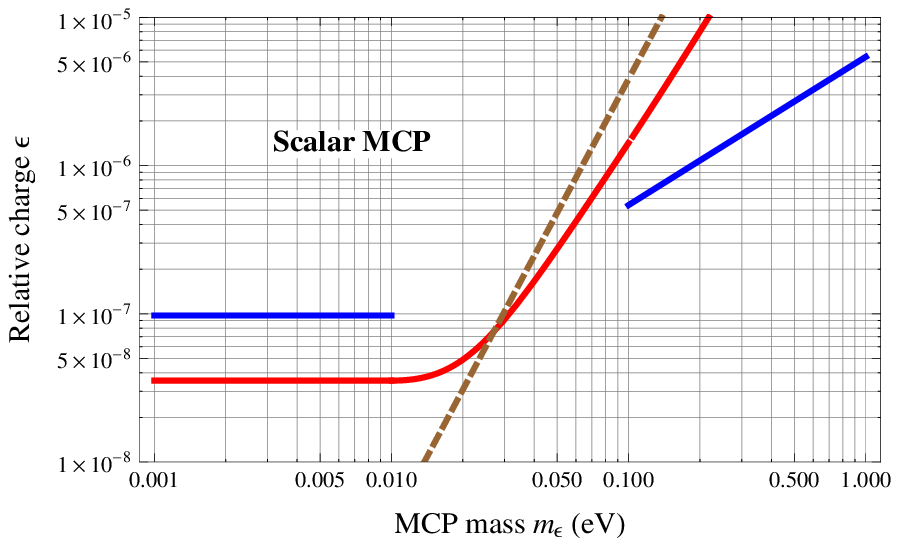}
\end{center}
\caption{Exclusion plots for MCP particles at 95\% c.l. deriving from the dichroism and birefringence values of Equations (\ref{BirefringencePVLAS}) and (\ref{DichroismPVLAS}). Top panel: Fermion MCP. Bottom panel: scalar MCP. The excluded region is above the curves. The limit derived from rotation dominates at small masses, whereas the limit of birefringence is effective at large masses. The two branches of the birefringence curve are not connected in the mass range around $\chi=1$ (dashed line), where $\Delta n$ changes sign. The two branches of the dichroism curve are joined by a cubic spline.}
\label{MCPsLimits}
\end{figure}

In Figure \ref{MCPsLimits} we show the exclusion plots on the existence of milli-charged particles. Two independent limits are derived from the birefringence and the dichroism measurements of Equations (\ref{BirefringencePVLAS}) and (\ref{DichroismPVLAS}), the latter being more stringent in the low-mass range ($m_\epsilon\leq0.1$~eV), whereas the former is dominating the high-mass range. We explicitly note that the Fermion exclusion plot applies also to all types of neutrinos, limiting their charge to be less than $\approx3\times10^{-8}e$ for mass smaller than 10~meV. 

\section{Conclusions}

We have presented a detailed report of the status of the PVLAS experiment, which strives to push further the frontier of the opto-magnetic polarimetry of small signals. As for the magnetic birefringence of vacuum, the new measurements are approaching the goal of the experiment. The measurements have given new limits also on the existence of hypothetical particles which couple to two photons, both axion-like and milli-charged. The sensitivity, although improved with respect to the past, has not yet reached the level that would guarantee the capability to perform the measurement in a reasonable time. The challenge of the experiment is now to lower the wide band noise. A few tests are ongoing, which should reduce the noise or at least shade light on its nature. Among them, we plan to rotate the magnets faster to reduce the incidence of the $1/f$ noise, to further reduce the scattered light, to search for mirrors with even higher finesse and smaller intrinsic birefringence, and to test the possibility to significantly lower the temperature of the mirrors. 

\section*{Acknowledgments}

We gratefully acknowledge the invaluable technical help and infinite patience of Luca Landi from the University of Ferrara.

\end{document}